\documentclass[12pt]{article}
\usepackage[a4paper,margin=1in]{geometry}
\usepackage[titletoc,title]{appendix}
\usepackage{changepage}
\usepackage[utf8x]{inputenc}
\usepackage{textcomp,marvosym}
\usepackage{fixltx2e}
\usepackage{amsmath,amssymb}
\usepackage{cite}
\usepackage{nameref,hyperref}
\usepackage{authblk}
\usepackage{algorithm}
\usepackage{algpseudocode}
\usepackage{dsfont}
\usepackage{xcolor}
\usepackage{enumitem}
\usepackage{caption,setspace}
\usepackage{subcaption}
\usepackage{pdflscape}
\usepackage{graphicx}
\newcommand{\dydx}[2]{\frac{\text{d} #1}{\text{d} #2}}

\renewcommand{\eqref}[1]{Equation~(\ref{#1})}
\newcommand{\Prob}[1]{\mathbb{P}(#1)}
\newcommand{\CondProb}[2]{\mathbb{P}(#1 \mid #2)}
\newcommand{\CondE}[2]{\mathbb{E}\left[#1 \mid #2\right]}
\newcommand{\CondC}[2]{\mathbb{C}\left[#1 \mid #2\right]}
\newcommand{\PDF}[1]{p(#1)}
\newcommand{\CondPDF}[2]{p(#1 \mid #2)}

\newcommand{\Kernel}[2]{q(#1 \mid #2)}
\newcommand{\E}[1]{\mathbb{E}\left[#1\right]}
\newcommand{\V}[1]{\text{Var}\left[#1\right]}

\newcommand{\bvec}[1]{\mathbf{#1}}
\newcommand{\ind}[2]{\mathds{1}_{#1}\left(#2\right)}

\DeclareMathOperator*{\diag}{diag}

\newcommand{\paramvec}{\boldsymbol{\theta}}

\newcommand{\paramspace}{\boldsymbol{\Theta}}
\newcommand{\discrep}[2]{\rho( #1, #2)}

\newcommand{\doi}[1]{DOI:\href{https://doi.org/#1}{#1}}

\title{Simulation and inference algorithms for stochastic biochemical reaction networks: from basic concepts to state-of-the-art}

\author[1]{David~J. Warne}
\author[2]{Ruth~E. Baker}
\author[1]{Matthew~J. Simpson\footnote{To whom correspondence should be addressed. E-mail: matthew.simpson@qut.edu.au}}

\affil[1]{School of Mathematical Sciences, Queensland University of Technology, Brisbane, Queensland 4001, Australia}
\affil[2]{Mathematical Institute, University of Oxford, Oxford, OX2 6GG, United Kingdom}

\begin{document}

\maketitle
\begin{abstract}
Stochasticity is a key characteristic of intracellular processes such as gene regulation and chemical signalling. Therefore, characterising stochastic effects in biochemical systems is essential to understand the complex dynamics of living things. Mathematical idealisations of biochemically reacting systems must be able to capture stochastic phenomena. While robust theory exists to describe such stochastic models, the computational challenges in exploring these models can be a significant burden in practice since realistic models are analytically intractable. Determining the expected behaviour and variability of a stochastic biochemical reaction network requires many probabilistic simulations of its evolution. Using a biochemical reaction network model to assist in the interpretation of time course data from a biological experiment is an even greater challenge due to the intractability of the likelihood function for determining observation probabilities. These computational challenges have been subjects of active research for over four decades. In this review, we present an accessible discussion of the major historical developments and state-of-the-art computational techniques relevant to simulation and inference problems for stochastic biochemical reaction network models. Detailed algorithms for particularly important methods are described and complemented with MATLAB$^\text{\textregistered}$ implementations. As a result, this review provides a practical and accessible introduction to computational methods for stochastic models within the life sciences community.

\end{abstract}


\section{Introduction}
Many biochemical processes within living cells, such as regulation of gene expression, are stochastic~\cite{Abkowitz1996,Arkin1998,Kaern2005,McAdams1997,Raj2008}; that is, randomness or noise is an essential component of living things. Internal and external factors are responsible for this randomness~\cite{Elowitz2002,Keren2015,Soltani2016,Taniguchi2010}, particularly within systems where low copy numbers of certain chemical species greatly affect the system dynamics~\cite{Fedoroff2002}. Intracellular stochastic effects are key components of normal cellular function~\cite{Eldar2010} and have a direct influence on the heterogeneity of multicellular organisms~\cite{Smith2018}. Furthermore, stochasticity of biochemical processes can play a role in the onset of disease~\cite{Feinberg2014,Gupta2011} and immune responses~\cite{Satija2014}. Stochastic phenomena, such as resonance~\cite{Moss1995}, focussing~\cite{Paulsson2000}, and bistability~\cite{Bressloff2017,Thattai2001,Tian2006}, are not captured by traditional deterministic chemical rate equation models. These stochastic effects must be captured by appropriate theoretical models. A standard approach is to consider a biochemical reaction network as a well-mixed population of molecules that diffuse, collide and react probabilistically. The stochastic law of mass action is invoked to determine the probabilities of reaction events over time~\cite{Gillespie1977,Wilkinson2012}. The resulting time-series of biochemical populations may be analysed to determine both the average behaviour and variability~\cite{Schnoerr2017}. This powerful approach to modelling biochemical kinetics can be extended to deal with more biologically realistic settings that include spatial heterogeneity with molecular populations being well-mixed only locally~\cite{Erban2009,Fange2010,Isaacson2008,Turner2004}. 

In practice, stochastic biochemical reaction network models are analytically intractable meaning that most approaches are entirely computational. Two distinct, yet related, computational problems are of particular importance: (i) the \textit{forwards} problem that deals with the simulation of the evolution of a biochemical reaction network forwards in time; and (ii) the \textit{inverse} problem that deals with the inference of unknown model parameters given time-course observations. Over the last four decades, significant attention has been given to these problems. Gillespie et~al.~\cite{Gillespie2013} describe the key algorithmic advances in the history of the forwards problem and Higham~\cite{Higham2008} provides an accessible introduction connecting stochastic approaches with deterministic counterparts. Recently, Schnoerr et~al.~\cite{Schnoerr2017}  provide a detailed review of the forwards problem with a focus on analytical methods. Golightly and Wilkinson~\cite{Golightly2011}, Toni et~al.~\cite{Toni2009} and Sunn{\aa}ker et~al.~\cite{Sunnaker2013} review techniques relevant to the inverse problem. 

Given the relevance of stochastic computational methods to the life sciences, the aim of this review is to present an accessible summary of computational aspects relating to efficient simulation for both the forwards and inverse problems.   
 Practical examples and algorithmic descriptions are presented and aimed at applied mathematicians and applied statisticians with interests in the life sciences. However, we expect the techniques presented here will also be of interest to the wider life sciences community. Supplementary material provides clearly documented code examples (available from GitHub \href{https://github.com/ProfMJSimpson/Warne2018}{https://github.com/ProfMJSimpson/Warne2018}) using the MATLAB$^\text{\textregistered}$ programming language.

\section{Biochemical reaction networks}
\label{sec:BCRN}
We provide an algorithmic introduction to stochastic biochemical reaction network models. In the literature, rigorous theory exists for these stochastic modelling approaches~\cite{Gillespie1992}. However, we focus on an informal definition useful for understanding computational methods in practice. Relevant theory on the chemical master equation, Markov processes and stochastic differential equations is not discussed in any detail (See Erban et~al.~\cite{Erban2007}, Higham~\cite{Higham2001} and Wilkinson~\cite{Wilkinson2012} for accessible introductions to these topics).   

Consider a domain, for example, a cell nucleus, that contains a number of \textit{chemical species}. The population count for a chemical species is a non-negative integer called its \textit{copy number}. A biochemical reaction network model is specified by a set of chemical formulae that determine how the chemical species interact. 
For example, $X + 2Y \rightarrow Z + W$, states ``one $X$ molecule and two $Y$ molecules react to produce one $Z$ molecule and one $W$ molecule''. If a chemical species is involved in a reaction, then the number of molecules required as reactants or produced as products are called \textit{stoichiometric coefficients}. In the example, $Y$ has a reactant stoichiometric coefficient of two, and $Z$ has a product stoichiometric coefficient of one.

\subsection{A computational definition}

Consider a set of $M$ reactions involving $N$ chemical species with copy numbers $X_1(t), \ldots, X_N(t)$ at time $t$. The state vector is an $N \times 1$ vector of copy numbers, $\bvec{X}(t) = [X_1(t), \ldots, X_N(t)]^T$.
This represents the state of the population of chemical species at time $t$.
When a reaction occurs, the copy numbers of the reactants and products are altered according to their respective stoichiometric coefficients. The net state change caused by a reaction event is called its stoichiometric vector. If reaction $j$ occurs, then a new state is obtained by adding its stoichiometric vector, $\boldsymbol{\nu}_j$, that is,
\begin{equation}
\label{eq:svup}
\bvec{X}(t) = \bvec{X}(t^-) + \boldsymbol{\nu}_j,
\end{equation}
where $t^-$ denotes the time immediately preceding the reaction event. 
The vectors $\boldsymbol{\nu}_1, \ldots, \boldsymbol{\nu}_M$ are obtained through $\boldsymbol{\nu}_j = \boldsymbol{\nu}_j^+ - \boldsymbol{\nu}_j^-$,
where $\boldsymbol{\nu}_j^-$ and $\boldsymbol{\nu}_j^+$ are, respectively, vectors of the reactant and product stoichiometric coefficients of the chemical formula of reaction $j$. \eqref{eq:svup} describes how reaction $j$  affects the system. 

Gillespie~\cite{Gillespie1977,Gillespie1992} presents the fundamental theoretical framework that provides a probabilistic rule for the occurrence of reaction events. We shall not focus on the details here, but the essential concept is based on the stochastic law of mass action. Informally,
\begin{equation}
\label{eq:lma}
\Prob{\text{Reaction }j\text{ occurs in } [t, t+\text{d}t) } \propto \text{d}t \times \text{\# of possible reactant combinations}.  
\end{equation}  
The tacit assumption is that the system is well-mixed with molecules equally likely to be found anywhere in the domain. The right hand side of \eqref{eq:lma} is typically expressed as $a_j(\bvec{X}(t))\text{d}t$, where $a_j(\bvec{X}(t))$ is the \textit{propensity function} of reaction $j$. That is,
\begin{equation}
\label{eq:propf}
a_j(\bvec{X}(t)) = \text{constant } \times \text{ total combinations in }\bvec{X}(t) \text{ for reaction }j,
\end{equation}
where the positive constant is known as the \textit{kinetic rate parameter}\footnote{These are not ``rates'' but scale factors on reaction event probabilities. A ``slow'' reaction with low kinetic rate may still occur rapidly, but the probability of this event is low.}. Equations~(\ref{eq:svup})--(\ref{eq:propf}) are the main concepts needed to consider computational methods for the forwards problem. Importantly, Equations~(\ref{eq:svup}) and (\ref{eq:lma}) indicate that the possible model states are discrete, but state changes occur in continuous time.

\subsection{Two examples}
We now provide some representative examples of biochemical reaction networks that will be used throughout this review. 

\subsubsection{Mono-molecular chain}
Consider two chemical species, $A$ and $B$, and three reactions that form a mono-molecular chain,
\begin{equation}
\label{eq:monomol}
\underbrace{\emptyset \overset{k_1}{\rightarrow} A}_{\substack{\text{external production}\\ \text{ of $A$ molecules}}}, \quad \underbrace{A \overset{k_2}{\rightarrow} B}_{\substack{\text{decay of $A$ molecules}\\ \text{into $B$ molecules}}}, \quad \underbrace{B \overset{k_3}{\rightarrow} \emptyset}_{\substack{\text{external consumption}\\ \text{of $B$ molecules}}},
\end{equation}
with kinetic rate parameters $k_1$, $k_2$, and $k_3$. We adopt the convention that $\emptyset$ indicates the reactions are part of an open system involving external chemical processes that are not explicitly represented in \eqref{eq:monomol}. Given the state vector, $\bvec{X}(t) = [A(t),B(t)]^T$, the respective propensity functions are
\begin{equation}
a_1(\bvec{X}(t)) = k_1, \quad a_2(\bvec{X}(t)) = k_2 A(t), \quad a_3(\bvec{X}(t)) = k_3 B(t),
\end{equation}
and the stoichiometric vectors are
\begin{equation}
\boldsymbol{\nu}_1 = \begin{bmatrix}
1 \\
0
\end{bmatrix}, \quad \boldsymbol{\nu}_2 = \begin{bmatrix}
-1 \\
1
\end{bmatrix}, \quad \boldsymbol{\nu}_3 = \begin{bmatrix}
0 \\
-1
\end{bmatrix}.
\end{equation}
This mono-molecular chain is interesting since it is a part of general class of biochemical reaction networks that are analytically tractable, though they are only applicable to relatively simple biochemical processes~\cite{Jahnke2007}. 

\subsubsection{Enzyme kinetics}
A biologically applicable biochemical reaction network describes the catalytic conversion of a substrate, $S$, into a product, $P$, via an enzymatic reaction involving enzyme, $E$. This is described by Michaelis--Menten enzyme kinetics~\cite{Michaelis1913,Rao2003},
\begin{equation}
\label{eq:michment}
\underbrace{E + S \overset{k_1}{\rightarrow} C}_{\substack{\text{ enzyme and substrate molecules} \\ \text{combine to form a complex}}}, \quad \underbrace{C \overset{k_2}{\rightarrow} E + S}_{\substack{\text{decay of a complex}}}, \quad \underbrace{C \overset{k_3}{\rightarrow} E+ P}_{\substack{\text{catalytic conversion}\\ \text{ of substrate to product}}},
\end{equation}
with kinetic rate parameters, $k_1$, $k_2$, and $k_3$. This particular enzyme kinetic model is a closed system. Here we have the state vector $\bvec{X}(t) = [E(t),S(t),C(t),P(t)]^T$, propensity functions
\begin{equation}
a_1(\bvec{X}(t)) = k_1E(t)S(t), \quad a_2(\bvec{X}(t)) = k_2 C(t), \quad a_3(\bvec{X}(t)) = k_3 C(t),
\end{equation}
and stoichiometric vectors
\begin{equation}
\boldsymbol{\nu}_1 = \begin{bmatrix}
-1 \\
-1 \\
1 \\
0
\end{bmatrix}, \quad \boldsymbol{\nu}_2 = \begin{bmatrix}
1 \\
1 \\
-1 \\
0
\end{bmatrix}, \quad \boldsymbol{\nu}_3 = \begin{bmatrix}
1 \\
0 \\
-1 \\
1
\end{bmatrix}.
\end{equation}
Since the first chemical formula involves two reactant molecules, there is significantly less progress that can be made without computational methods. 

See \texttt{MonoMolecularChain.m} and \texttt{MichaelisMenten.m} for example code to generate useful data structures for these biochemical reaction networks.
These two biochemical reaction networks have been selected to demonstrate two stereotypical problems. In the first instance, the mono-molecular chain model, the network structure enables progress to be made analytically. The enzyme kinetic model, however, represents a more realistic case in which  computational methods are required. The focus on these two representative models is done so that the exposition is clear.

\section{The forwards problem}
\label{sec:sim}
Given a biochemical reaction network with known kinetic rate parameters and some initial state vector, $\bvec{X}(0)~=~\bvec{x}_0$, we consider the forwards problem. That is, we wish to predict the future evolution. Since we are dealing with stochastic models, all our methods for dealing with future predictions will involve probabilities, random numbers and uncertainty. We rely on standard code libraries\footnote{We utilise the Statistics and Machine Learning Toolbox within the MATLAB$^{\text{\textregistered}}$ environment for generating random samples from any of the standard probability distributions.} for generating samples from uniform (i.e., all outcomes equally likely), Gaussian (i.e., the bell curve), exponential (i.e., time between events), and Poisson distributions (i.e., number of events over a time interval) and do not discuss algorithms for generating samples from these distributions. This enables the focus of this review to be on algorithms specific to biochemical reaction networks.

There are two key aspects to the forwards problem: (i) the simulation of a biochemical reaction network that replicates the random reaction events over time; and (ii) the calculation of average behaviour and variability among all possible sequences of reaction events. Relevant algorithms for dealing with both these aspects are reviewed and demonstrated.

\subsection{Generation of sample paths}
Here, we consider algorithms that deal with the simulation of biochemical reaction network evolution. These algorithms are probabilistic, that is, the output of no two simulations, called \textit{sample paths,} of the same biochemical reaction network will be identical. The most fundamental stochastic simulation algorithms (SSAs) for sample path generation are based on the work of Gillespie~\cite{Gillespie1977,Gillespie2000,Gillespie2001}, Gibson and Bruck~\cite{Gibson2000}, and Anderson~\cite{Anderson2007}.

\subsubsection{Exact stochastic simulation algorithms}
\label{sec:essa}
Exact SSAs generate sample paths, over some interval $t \in [0, T]$, that identically follow the probability laws of the fundamental theory of stochastic chemical kinetics~\cite{Gillespie1992}. 
Take a sufficiently small time interval, $[t,t+\text{d} t)$, such that the probability of multiple reactions occurring in this interval is zero. In such a case, the reactions are mutually exclusive events. Hence, based on Equations~(\ref{eq:lma})~and~(\ref{eq:propf}), the probability of any reaction event occurring in $[t,t+\text{d}t)$ is the sum of the individual event probabilities,
\begin{align*}
\Prob{\text{Any reaction occurs in } [t, t+\text{d}t) } &= \Prob{\text{Reaction }1\text{ occurs in } [t, t+\text{d}t) } + \cdots \\ &\quad+\Prob{\text{Reaction }M\text{ occurs in } [t, t+\text{d}t) }\\
&= a_0(\bvec{X}(t))\text{d}t,
\end{align*}
where $a_0(\bvec{X}(t)) =  a_1(\bvec{X}(t)) + \cdots + a_M(\bvec{X}(t))$ is the total reaction propensity function. 
Therefore, if we know that the next reaction occurs at time $s \in [t, t+\text{d}t)$, then we can: (i) randomly select a reaction with probabilities, $a_1(\bvec{X}(s))/a_0(\bvec{X}(s)), \ldots, a_M(\bvec{X}(s))/a_0(\bvec{X}(s))$; and (ii) update the state vector according to the respective stoichiometric vector, $\boldsymbol{\nu}_1,  \ldots, \boldsymbol{\nu}_M$. 

All that remains for an exact simulation method is to determine the time of the next reaction event. Gillespie~\cite{Gillespie1977} demonstrates that the time interval between reactions, $\Delta t$, may be treated as a random variable that is exponentially distributed with rate $a_0(\bvec{X}(t))$, that is, $\Delta t \sim \text{Exp}(a_0(\bvec{X}(t)))$.
Therefore, we arrive at the most fundamental exact SSA, the \emph{Gillespie direct method}:
\begin{enumerate}
	\item initialise, time $t = 0$ and state vector $\bvec{X} = \bvec{x}_0$;
	\item calculate propensities, $a_1(\bvec{X}), \ldots, a_M(\bvec{X})$, and  $a_0(\bvec{X}) = a_1(\bvec{X}) +\cdots + a_M(\bvec{X})$;\label{line:l1}
	\item generate random sample of the time to next reaction, $\Delta t \sim \text{Exp}(a_0(\bvec{X}))$;
	\item if $t + \Delta t > T$, then terminate the simulation, otherwise, go to step \ref{line:sample};
	\item randomly select integer $j$ from the set $\left\{1,\ldots, M\right\}$ with\\ $\Prob{j= 1} =a_1(\bvec{X})/a_0(\bvec{X})$, $\ldots, \Prob{j=M} = a_M(\bvec{X})/a_0(\bvec{X})$;\label{line:sample}
	\item update state, $\bvec{X} = \bvec{X} + \boldsymbol{\nu}_j$, and time $t = t + \Delta t$, then go to step \ref{line:l1}.  
\end{enumerate}
An example implementation, \texttt{GillespieDirectMethod.m}, and example usage, \texttt{DemoGillespie.m} are provided. Figure~\ref{fig:Sim} demonstrates sample paths generated by Gillespie direct method for the mono-molecular chain model (Figure~\ref{fig:Sim}(a)) and the enzyme kinetic model (Figure~\ref{fig:Sim}(b)).

A mathematically equivalent, but more computationally efficient, exact SSA formulation is derived by Gibson and Bruck~\cite{Gibson2000}. Their method independently tracks the next reaction time of each reaction separately. The next reaction to occur is the one with the smallest next reaction time, therefore no random selection of reaction events is required. It should, however, be noted that the Gillespie direct method may also be improved to yield the \emph{optimised direct method}~\cite{Cao2004} with similar performance benefits. Anderson~\cite{Anderson2007} further refines the method of Gibson and Bruck by scaling the times of each reaction so that the scaled times between reactions follow unit-rate exponential random variables. This scaling allows the method to be applied to more complex biochemical reaction networks with time-dependent propensity functions, however, the recently proposed Extrande method~\cite{Voliotis2016} is computationally superior. In Anderson's approach, scaled times are tracked for each reaction independently with $t_j$ being the current time at the natural scale of reaction $j$. This results in the \emph{modified next reaction method}:
\begin{enumerate}
	\item initialise, global time $t = 0$, state vector $\bvec{X} = \bvec{x}_0$ and scaled times $t_1 = t_2  = \cdots = t_M = 0$; 
	\item generate $M$ first reaction times, $s_1, \ldots, s_M \sim \text{Exp}(1)$;
	\item calculate propensities, $a_1(\bvec{X}), \ldots, a_M(\bvec{X})$\label{line:prop};
	\item rescale time to next reaction, $\Delta t_j = (s_j - t_j)/a_j(\bvec{X})$ for $j = 1,2, \ldots, M$;
	\item choose reaction $k$, such that $\Delta t_k = \min\left\{\Delta t_1, \ldots, \Delta t_M\right\}$; 
	\item if $t + \Delta t_k > T$ terminate simulation, otherwise go to step~\ref{line:update};
	\item update rescaled  times, $t_j = t_j + a_j(\bvec{X})\Delta t_k$ for $j = 1, \ldots, M$, state $\bvec{X} = \bvec{X} + \boldsymbol{\nu}_k$ and global time $t = t + \Delta t_k$\label{line:update};
	\item generate scaled next reaction time for reaction $k$, $\Delta s_k \sim \text{Exp}(1)$;
	\item update next scaled reaction time $s_k = s_k + \Delta s_k$, and go to step \ref{line:prop}.
\end{enumerate} 
An example implementation, \texttt{ModifiedNextReactionMethod.m}, and example usage, \texttt{DemoMNRM.m} are provided. Figure~\ref{fig:Sim} demonstrates sample paths generated by the modified next reaction method for the mono-molecular chain model (Figure~\ref{fig:Sim}(c)) and the enzyme kinetic model (Figure~\ref{fig:Sim}(d)). Note that the sample paths are different to those generated using the Gillespie direct method, despite the random number generators being initialised the same way. However, both represent exact sample paths, that is, sample paths that exactly follow the dynamics of the biochemical reaction network.

While the Gillespie direct method and the more efficient modified next reaction method and optimised direct method represent the most fundamental examples of exact SSAs, other advanced methods are also available to further improve the computational performance for large and complex biochemical reaction networks. Particular techniques include partial-propensity factorisation~\cite{Indurkhya2010}, rejection-based methods~\cite{Thanh2014,Thanh2017} and composition-rejection~\cite{Slepoy2008} methods. We do not discuss these approaches, but we highlight them to indicate that efficient exact SSA method development is still an active area of research.

\subsubsection{Approximate stochastic simulation algorithms}
Despite some computational improvements provided by the modified next reaction method~\cite{Anderson2007,Gibson2000}, all exact SSAs are computationally intractable for large biochemical populations and with many reactions, since every reaction event is simulated. Several approximate SSAs have been introduced in an attempt to reduce the computational burden while sacrificing accuracy.

\begin{landscape}
	\begin{figure}
		\centering
		\includegraphics[width=\linewidth]{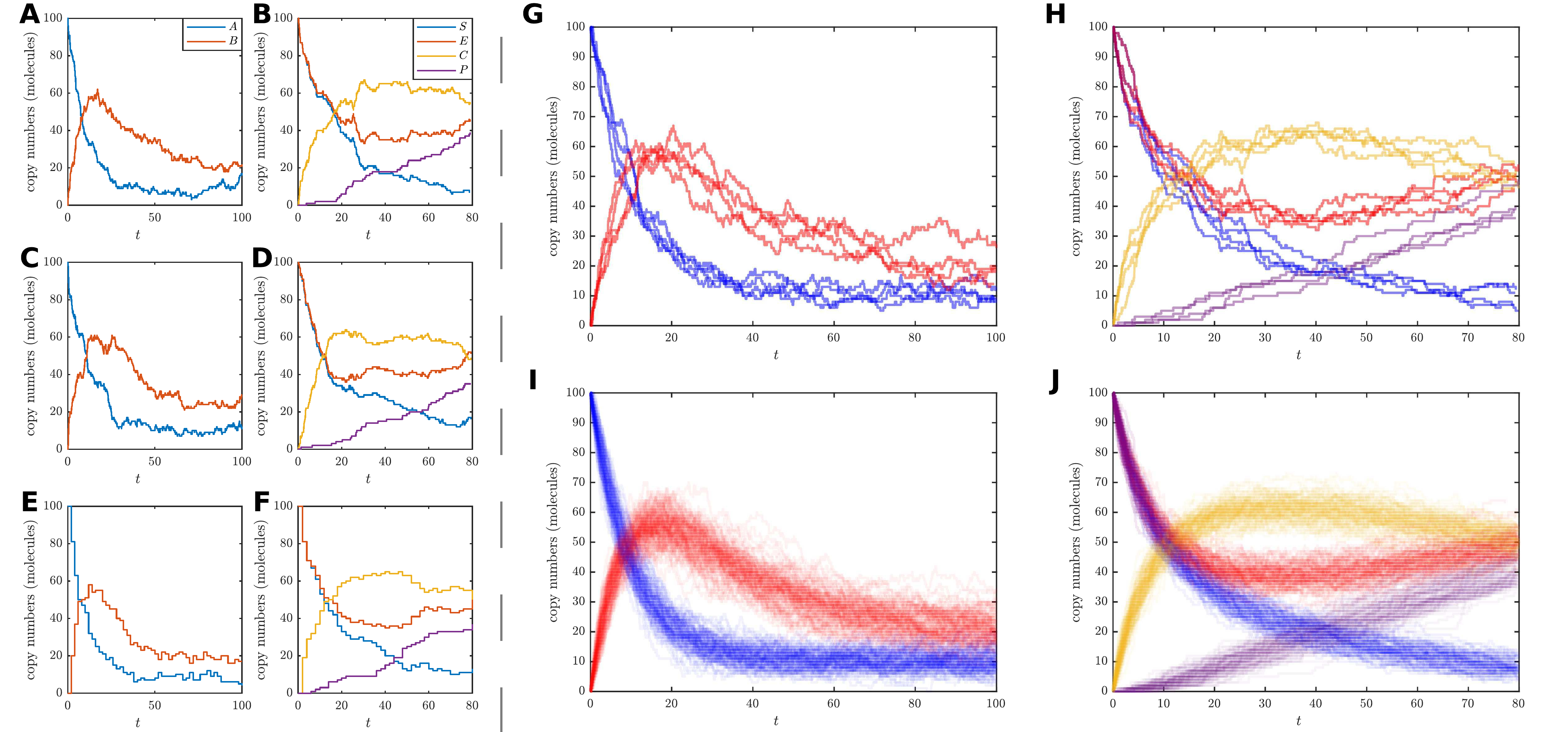}
		\caption{Examples of exact sample paths of the mono-molecular chain model using the (a) Gillespie direct method and (c) modified next reaction method; similarly exact sample paths of the enzyme kinetics model using the (b) Gillespie direct method and (d) modified next reaction method. Approximate sample paths may be computed with less computational burden using the tau leaping method with $\tau = 2$, at the expense of accuracy: (e) the mono-molecular chain model and (f) the enzyme kinetics model. Every sample path will be different; as demonstrated by four distinct simulations of (g) the mono-molecular chain model and (h) the enzyme kinetics model. However, trends are revealed when 100 simulations are overlaid to reveal states of higher probability density using (i) the mono-molecular chain model and (j) the enzyme kinetics model. The mono-molecular chain model simulations are configured with parameters $k_1 = 1.0$, $k_2 = 0.1$, $k_3 = 0.05$, and initial state $A(0) = 100$, $B(0) = 0$. The enzyme kinetics model simulations are configured with parameters $k_1 = 0.001$, $k_2 = 0.005$, $k_3 = 0.01$, and initial state $E(0) = 100$, $S(0) = 100$, $C(0) = 0$, $P(0) = 0$.} 
		\label{fig:Sim}
	\end{figure}
\end{landscape}

The main approximate SSA we consider is also developed by Gillespie~\cite{Gillespie2001} almost 25 years after the development of the Gillespie direct method. The key idea is to evolve the system in discrete time steps of length $\tau$, hold the propensity functions constant over the time interval $[t, t + \tau)$ and count the number of reaction events that occur. The state vector is then updated based on the net effect of all the reaction events. The number of reaction events within the interval can be shown to be a random variable distributed according to a Poisson distribution with mean $a_j(\bvec{X}(t))\tau$. If $Y_j$ denotes the number of reaction $j$ events in $[t, t+ \tau)$ then $Y_j \sim \text{Po}(a_j(\bvec{X}(t))\tau)$. The result is the \emph{tau leaping method}:
\begin{enumerate}
	\item initialise, time $t = 0$ and state $\bvec{Z} = \bvec{x}_0$;
	\item if $t + \tau > T$ then terminate simulation, otherwise continue\label{line:repeat};
	\item calculate propensities, $a_1(\bvec{Z}), \ldots, a_M(\bvec{Z})$;
	\item generate reaction event counts, $Y_j \sim \text{Po}(a_j(\bvec{Z})\tau)$ for $j = 1, \ldots, M$;
	\item update state, $\bvec{Z} = \bvec{Z} + Y_1 \boldsymbol{\nu}_1 + \cdots Y_M \boldsymbol{\nu}_M$, and time $t = t + \tau$;
	\item go to step \ref{line:repeat}.
\end{enumerate}
Note that we use the notation $\bvec{Z}(t)$ to denote an approximation of the true state $\bvec{X}(t)$. An example implementation, \texttt{TauLeapingMethod.m}, and example usage, \texttt{DemoTauLeap.m} are provided. Figure~\ref{fig:Sim} demonstrates sample paths generated by the tau leaping method for the mono-molecular chain model (Figure~\ref{fig:Sim}(e)) and the enzyme kinetic model (Figure~\ref{fig:Sim}(f)). Note that there is a visually obvious difference in the noise patterns of the tau leaping method sample paths and the exact SSA sample paths (Figure~\ref{fig:Sim}(a)--(d)).

The tau leaping method is the only approximate SSA that we will explicitly discuss as it captures the essence of what approximations try to achieve; trading accuracy for improved performance.  Several variations of the tau leaping method have been proposed to improve accuracy, such as adaptive $\tau$ selection~\cite{Cao2005,Cao2006}, implicit schemes~\cite{Rathinam2003} and the replacement of Poisson random variables with binomial random variables~\cite{Tian2004}. Hybrid methods that combine exact SSAs and approximations that split reactions into different time-scales are also particularly effective for large scale networks with reactions occurring on very different time-scales~\cite{Cao2005b,E2005}. Other approximate simulation approaches are, for example, based on a continuous state chemical Langevin equation approximation~\cite{Cotter2016,Gillespie2000,Wilkinson2009} and employ numerical schemes for stochastic differential equations~\cite{Burrage2003,Higham2001}.

\subsection{Computation of summary statistics}
Due to the stochastic nature of biochemical reaction networks, one cannot predict with certainty the future state of the system. Averaging over $n$ sample paths can, however, provide insights into likely future states. Figure~\ref{fig:Sim}(g),(h) show that there is considerable variation in four independent sample paths, $n = 4$, of the mono-molecular chain and enzyme kinetic models. However, there is still a qualitative similarity between them. This becomes more evident for $n = 100$ sample paths, as in Figure~\ref{fig:Sim}(i),(j). The natural extension is to consider average behaviour as $n \to \infty$. 

\subsubsection{Using the chemical master equation}
From a probability theory perspective, a biochemical reaction network model is a discrete-state, continuous-time Markov process. One key result for discrete-state, continuous-time Markov processes is, given an initial state, $\bvec{X}(t) = \bvec{x}_0$, one can describe how the probability distribution of states evolves. This is given by the chemical master equation~\cite{Gillespie1992}\footnote{In the theory of Markov processes, this equation is known as the \emph{Kolmogorov forward equation}.},
\begin{equation}
\label{eq:cme}
	\dydx{P(\bvec{x},t \mid \bvec{x}_0)}{t} = \underbrace{\sum_{j=1}^M a_j(\bvec{x} - \boldsymbol{\nu}_j)P(\bvec{x} - \boldsymbol{\nu}_j, t\mid \bvec{x}_0)}_{\substack{\text{probability increase from events}\\\text{that cause state change to }\bvec{x}}}\underbrace{ - P(\bvec{x},t \mid \bvec{x}_0 ) \sum_{j=1}^M a_j(\bvec{x})}_{\substack{\text{probability decrease from events}\\\text{that cause state change from }\bvec{x}}},
\end{equation} 
where $P(\bvec{x},t \mid \bvec{x}_0)$ is the probability that $\bvec{X}(t) = \bvec{x}$ given $\bvec{X}(0) = \bvec{x}_0$. Solving the chemical master equation provides an explicit means of computing the probability of being in any state at any time. Unfortunately, solutions to the chemical master equation are only known for special cases~\cite{Gadgil2005,Jahnke2007}.

\begin{figure}[h]
		\centering
		\includegraphics[width=\linewidth]{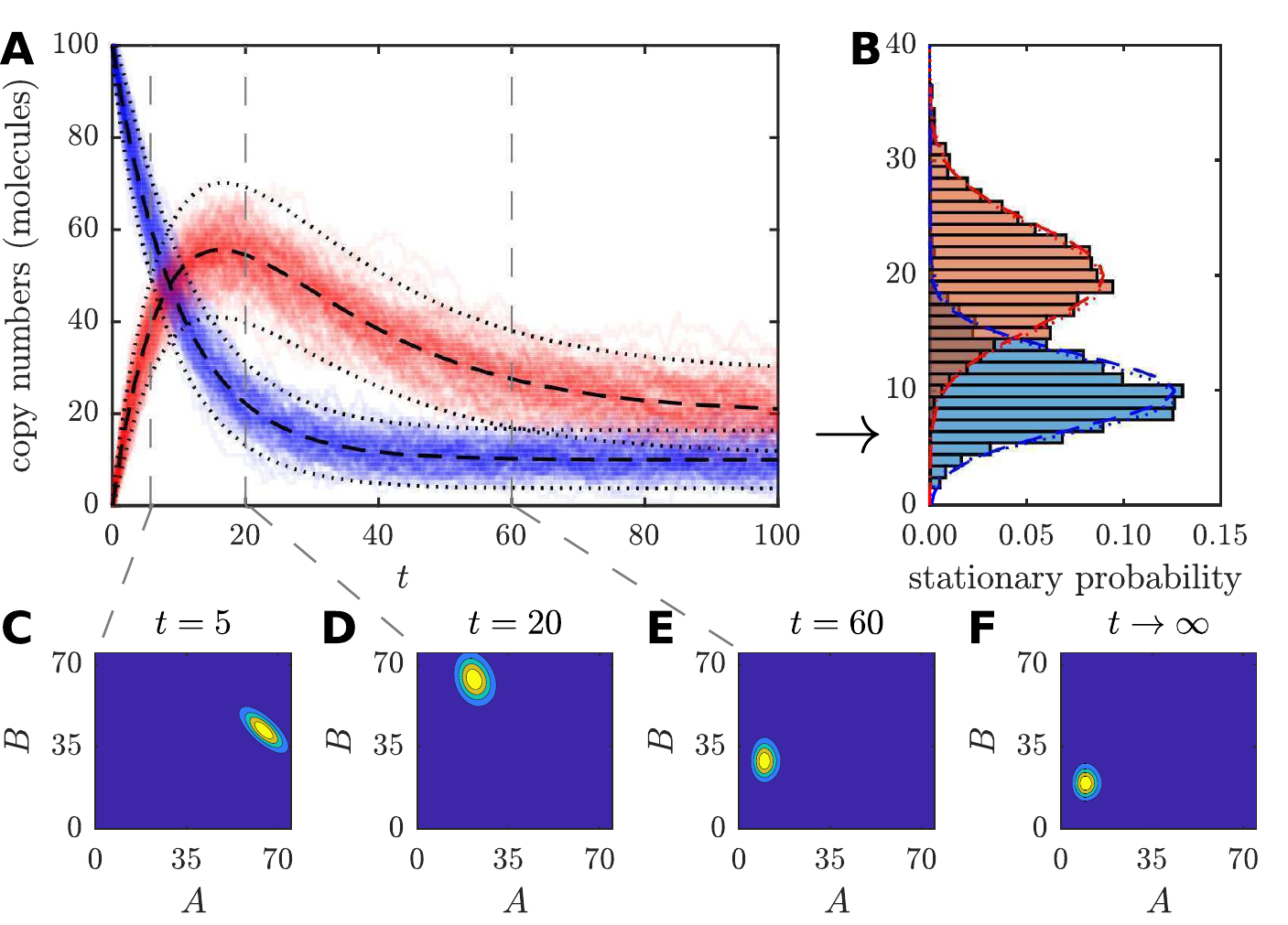}
		\caption{(a) The chemical master equation mean (black dashed) $\pm$ two standard deviations (black dots) of copy numbers of $A$ (blue) and $B$ (red) chemical species are displayed over simulated sample paths to demonstrate agreement. (b) The stationary distributions of $A$ and $B$ computed using: long running, $T = 1000$, simulated sample paths (blue/red histograms); Gaussian approximation (blue/red dashed) using long time limits of chemical master equation mean and variances; and the long time limit of the full chemical master equation solution (blue/red dots). Transient chemical master equation solution at times (c) $t = 5$, (d) $t = 20$, and (e) $t = 60$. (f) chemical master equation solution stationary distribution. The parameters are $k_1 = 1.0$, $k_2 = 0.1$, $k_3 = 0.05$, and initial state is $A(0) = 100$, $B(0) = 0$.}
		\label{fig:sumStatCME}
	\end{figure}

However, the mean and variance of the biochemical reaction network molecule copy numbers can sometimes be derived without solving the chemical master equation. For example, for the mono-molecular chain model (\eqref{eq:monomol}), one may use \eqref{eq:cme} to derive the following system of ordinary differential equations (see Appendix~\ref{sec:app_dev_cme}),
\begin{align}
\dydx{M_a(t)}{t} &= k_1 - k_2 M_a(t),\label{eq:Ma} \\ 
\dydx{M_b(t)}{t} &= k_2 M_a(t) - k_3 M_b (t),\label{eq:Mb} \\
\dydx{V_a(t)}{t} &= k_1 + k_2 M_a(t) - 2k_2 V_a(t),\label{eq:Va} \\ 
\dydx{V_b(t)}{t} &= k_2 M_a(t) + k_3 M_b(t) + 2 k_2 C_{a,b}(t) - k_3V_b(t),\label{eq:Vb} \\
\dydx{C_{a,b}(t)}{t} &= k_2 V_a(t) - k_2 M_a(t) - (k_2 + k_3)C_{a,b}(t), \label{eq:Cab}
\end{align}   
where $M_a(t)$ and $V_a(t)$ ($M_b(t)$ and $V_b(t)$) are the mean and variance of the copy number $A(t)$ ($B(t)$) over all possible sample paths. $C_{a,b}(t)$ is the covariance of between $A(t)$ and $B(t)$. Equations~(\ref{eq:Ma})--(\ref{eq:Cab}) are linear ordinary differential equations than can be solved analytically and the solution is plotted in Figure~\ref{fig:sumStatCME}(a) superimposed with a population of sample paths.

The long time limit behaviour of a biochemical reaction network can also be determined. For the mono-molecular chain, as $t \to \infty$ we have: $M_a(t) \to k_1/k_2$; $V_a(t) \to k_1/k_2$; $M_b(t) \to k_1/k_3$; $V_b(t) \to k_1/k_3$; and $C_{a,b}(t) \to 0$. We can approximate the long time steady state behaviour, called the \textit{stationary distribution}, of the mono-molecular chain model as two independent Gaussian random variables. That is, as $t \to \infty$, $A(t) \to A_\infty$ with $A_\infty \sim \mathcal{N}(k_1/k_2,k_1/k_2)$. Similarly, $B(t) \to B_\infty$ with $B_\infty \sim \mathcal{N}(k_1/k_3, k_1/k_3)$. This approximation is shown in Figure~\ref{fig:sumStatCME}(b) against histograms of sample paths generated using large values of $T$ (see example code, \texttt{DemoStatDist.m}).

In the case of the mono-molecular chain model, the chemical master equation is analytically tractable~\cite{Gadgil2005,Jahnke2007}. However, the solution is algebraically complicated and non-trivial to evaluate (see Appendix~\ref{sec:app_cme_comp}). The full chemical master equation solution in Figure~\ref{fig:sumStatCME}(c)--(e) and the true stationary distribution of two independent Poisson distributions is shown in Figure~\ref{fig:sumStatCME}(f). The true stationary distribution is also compared with the Gaussian approximation in Figure~\ref{fig:sumStatCME}(b); the approximation is reasonably accurate, however, we could have also reasonably surmised the true stationary distribution by noting that, for a Poisson distribution, the mean is equal to the variance. 
 
\subsubsection{Monte Carlo methods}

The chemical master equation can yield insight for special cases, however, for practical problems, such as the enzyme kinetic model, the chemical master equation is intractable and numerical methods are required. Here, we consider numerical estimation of the mean state vector at a fixed time, $T$.  

In probability theory, the mean of a distribution is defined via an expectation, 
\begin{equation}
\label{eq:expect}
\E{\bvec{X}(T)} = \sum_{\bvec{x} \in \Omega} \bvec{x}P(\bvec{x},T \mid \bvec{x}_0), 
\end{equation} 
where $\Omega$ is the set of all possible states. It is important to note that the methods we describe here are equally valid for a more general expectations of the form $\E{f(\bvec{X}(T))}$ where $f$ is some function that satisfies certain conditions~\cite{Giles2015}. 

We usually cannot compute \eqref{eq:expect} directly since $\Omega$ is typically infinite and the chemical master equation is intractable. However, exact SSAs provide methods for sampling the chemical master equation distribution, $\bvec{X}(T) \sim P(\bvec{x},T \mid \bvec{x}_0)$. This leads to the Monte Carlo estimator
\begin{equation}
\label{eq:mc}
\E{\bvec{X}(T)} \approx \hat{\bvec{X}}(T) = \frac{1}{n}\sum_{i=1}^n \bvec{X}(T)^{(i)},
\end{equation}
where $\bvec{X}(T)^{(1)}, \ldots, \bvec{X}(T)^{(n)}$ are $n$ independent sample paths of the biochemical reaction network of interest (see example implementation \texttt{MonteCarlo.m}). 

Unlike \eqref{eq:expect}, the Monte Carlo estimates, such as \eqref{eq:mc}, are random variables for finite $n$. This incurs a probabilistic error. A common measure of the accuracy of a Monte Carlo estimator, $\hat{\mu}(T)$, of some expectation, $\E{\mu(T)}$, is the \emph{mean-square error} that evaluates the average error behaviour and may be decomposed as follows,
\begin{equation}
\underbrace{\E{\left(\hat{\mu}(T) - \E{\mu(T)}\right)^2}}_{\text{ mean-square error}} = \underbrace{\V{\hat{\mu}(T)}}_{\text{Estimator variance}} + \left(\underbrace{\E{\hat{\mu}(T)} - \E{\mu(T)}}_{\text{Estimator bias}}\right)^2.\label{eq:MSE}
\end{equation}
\eqref{eq:MSE} highlights that there are two sources of error in a Monte Carlo estimator, the estimator \textit{variance} and \textit{bias}, and much of the discussion that follows deals with how to balance both these error sources in a computationally efficient manner. 

Through analysis of the mean-square error of an estimator, the rate at which the mean-square error decays as $n$ increases can be determined. Hence, we can determine how large $n$ needs to be satisfy the condition
\begin{equation}
\sqrt{\E{\left(\hat{\mu}(T) - \E{\mu(T)}\right)^2}} \leq ch, \label{eq:MSEerrortol}
\end{equation} 
where $c$ is a positive constant and $h$ is called the \emph{error tolerance}.

Since $\E{\hat{\bvec{X}}(T)} = \E{\bvec{X}(T)}$, the bias term in \eqref{eq:MSE} is zero and we call $\hat{\bvec{X}}(T)$ an \textit{unbiased} estimator of $\E{\bvec{X}(T)}$. For an unbiased estimator, the mean-square error is equal to the estimator variance. Furthermore, $\V{\hat{\bvec{X}}(T)} = \V{\bvec{X}(T)}/n$, so the estimator variance decreases linearly with $n$, for sufficiently large $n$. Therefore, $h \propto 1/\sqrt{n}$. That is, to halve $h$, one must increase $n$ by a factor of four. This may be prohibitive with exact SSAs, especially for biochemical reaction networks with large variance. In the context of the Monte Carlo estimator using an exact SSA (\eqref{eq:mc}), for $n$ sufficiently large, the central limit theorem (CLT) states that $\hat{\bvec{X}}(T) \sim \mathcal{N}\left(\E{\bvec{X}(T)},\V{\bvec{X}(T)}/n\right)$ (see Wilkinson~\cite{Wilkinson2012} for a good discussion on the CLT).

Computational improvements can be achieved by using an approximate SSA, such as the tau leaping method,
\begin{equation}
\label{eq:tlmmc}
\E{\bvec{X}(T)} \approx \E{\bvec{Z}(T)} \approx \hat{\bvec{Z}}(T) = \frac{1}{n}\sum_{i=1}^n \bvec{Z}(T)^{(i)},
\end{equation}
where $\bvec{Z}(T)^{(1)}, \ldots, \bvec{Z}(T)^{(n)}$ are $n$ independent approximate sample paths of the biochemical reaction network of interest (see the example implementation \texttt{MonteCarloTauLeap.m}). 

\begin{figure}[h]
	\centering
	\includegraphics[width=\linewidth]{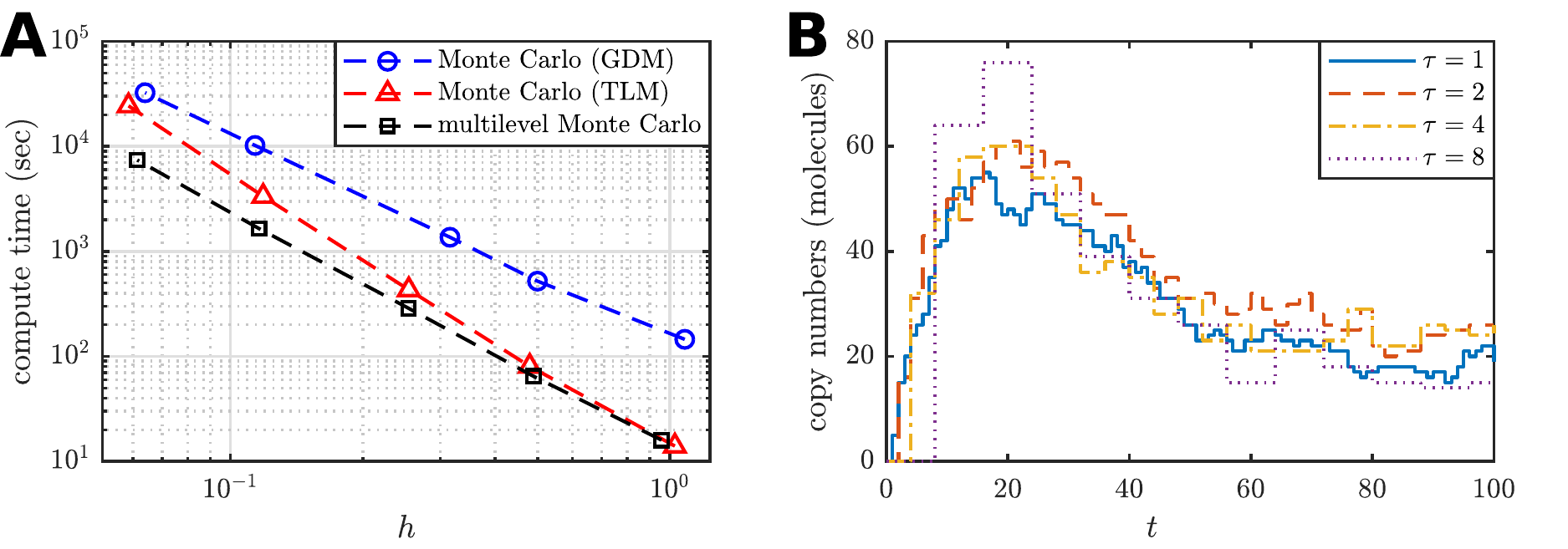}
	\caption{ (a) Improved performance from MLMC when estimating $\E{B(T)}$ at $T = 100$ using the mono-molecular chain model with parameters $k_1 = 10$, $k_2 = 0.1$, $k_3 = 0.5$, and initial condition $A(0) = 1000$, $B(0) = 0$; the computational advantage of the tau leaping method (red triangles dashed) over the Gillespie direct method (blue circles dashed) for Monte Carlo diminishes as the required error tolerance decreases. The MLMC method (black squares dashed) exploits the correlated tau leaping method to obtain sustained computational efficiency. (b) Demonstration of correlated tau leaping method simulations for nested $\tau$ steps; sample paths using a step of $\tau = 2$ (red dashed), $\tau = 4$ (yellow dot-dashed), and $\tau = 8$ (purple dots) are all correlated with the same $\tau = 1$ trajectory (blue solid). Computations are performed using an Intel$^\text{\textregistered}$ Core\texttrademark~i7-5600U CPU (2.6 GHz).} 
	\label{fig:MLMC}
\end{figure}

 Note that, $\E{\hat{\bvec{Z}}(T)} = \E{\bvec{Z}(T)}$. Since, $\E{\bvec{Z}(T)} \neq \E{\bvec{X}(T)}$ in general, we call $\hat{\bvec{Z}(T)}$ a \textit{biased} estimator. Even in the limit of $n \to \infty$, the bias term in \eqref{eq:MSE} may not be zero, which incurs a lower bound on the best achievable error tolerance, $h$, for fixed $\tau$. However, it has been shown that the bias of the tau leaping method decays linearly with $\tau$~\cite{Anderson2011,Li2007}. Therefore, to satisfy the error tolerance condition (\eqref{eq:MSEerrortol}) we not only require $n \propto 1/h^2$, but also $\tau \propto h$. That is, as $h$ decrease the performance improvement of Monte Carlo with the tau leaping method reduces by a factor of $\tau$, because the computational cost of the tau leaping method is proportional to $1/\tau$. In Figure~\ref{fig:MLMC}(a), the decay of the computational advantage in using the tau leaping method for Monte Carlo over the Gillespie direct method is evident, and eventually the tau leaping method will be more computationally burdensome than the Gillespie direct method. By the CLT, for large $n$, we have $\hat{\bvec{Z}}(T) \sim \mathcal{N}(\E{\bvec{Z}(T)},\V{\bvec{Z}(T)}/n)$. 

The utility of the tau leaping method for accurate (or exact) Monte Carlo estimation is identified by Anderson and Higham~\cite{Anderson2012} through extending the idea of multilevel Monte Carlo (MLMC) originally proposed by Giles for stochastic differential equations~\cite{Giles2008,Giles2015}. Consider a sequence of $L+1$ tau leaping method time steps $\tau_0, \tau_1,\ldots, \tau_L$, with $\tau_\ell < \tau_{\ell-1}$ for $\ell = 1, \ldots, L$. Let $\bvec{Z}_\ell(T)$ denote the state vector of a tau leaping method approximation using time step $\tau_\ell$. Assume $\tau_L$ is small enough that $\E{\bvec{Z}_L(T)}$ is a good approximation of $\E{\bvec{X}(T)}$. Note, for large $\tau_\ell$ (small $\ell$), sample paths are cheap to generate, but inaccurate; conversely, small $\tau_\ell$ (large $\ell$) results in computationally expensive, but accurate sample paths. 

We can write
\begin{align}
\label{eq:mlmc}
\E{\bvec{X}(T)} &\approx \underbrace{\E{\bvec{Z}_L(T)}}_{\substack{\text{low bias}\\\text{ approximation}}} \notag\\
 &= \underbrace{\E{\bvec{Z}_{L-1}(T)}}_{\substack{\text{slightly biased}\\\text{ approximation}}} + \underbrace{\E{\bvec{Z}_L(T) - \bvec{Z}_{L-1}(T)}}_{\substack{\text{bias correction}}} \notag\\
&= \underbrace{\E{\bvec{Z}_{L-2}(T)}}_{\substack{\text{slightly more biased}\\\text{ approximation}}} + \underbrace{\E{\bvec{Z}_{L-1}(T) - \bvec{Z}_{L-2}(T)} + \E{\bvec{Z}_L(T) - \bvec{Z}_{L-1}(T)}}_{\substack{\text{two bias corrections}}} \notag\\
&\,\,\,\vdots\notag\\
&= \underbrace{\E{\bvec{Z}_{0}(T)}}_{\substack{\text{very biased}\\\text{ approximation}}} + \underbrace{\sum_{\ell=1}^{L}\E{\bvec{Z}_{\ell}(T) - \bvec{Z}_{\ell-1}(T)}}_{\substack{L\text{ bias corrections}}}.
\end{align} 
Importantly, the final estimator on the right of \eqref{eq:mlmc}, called the multilevel telescoping summation, is equivalent in bias to $\E{\bvec{Z}_L(T)}$. At first glance, \eqref{eq:mlmc} looks to have complicated the computational problem and inevitably decreased performance of the Monte Carlo estimator. The insight of Giles~\cite{Giles2008}, in the context of stochastic differential equation models for financial applications, is that the bias correction terms may be computed using Monte Carlo approaches that involve generating highly correlated sample paths in estimation of each of the correction terms, thus reducing the variance in the bias corrections.  If the correlation is strong enough, then the variance decays such that few of the most accurate tau leaping method sample paths are required; this can result in significant computational savings.

A contribution of Anderson and Higham~\cite{Anderson2012} is an efficient method of generating correlated tau leaping method sample path pairs $(\bvec{Z}_{\ell}(T),\bvec{Z}_{\ell-1}(T))$ in the case when $\tau_{\ell} = \tau_{\ell-1}/\delta$ for some positive integer scale factor $\delta$. The algorithm is based on the property that the sum of two Poisson random variables is also a Poisson random variable. This enables the sample path with $\tau_{\ell-1}$ to be constructed as an approximation to the sample path with $\tau_{\ell}$ directly. Figure~\ref{fig:MLMC}(b) demonstrates tau leaping method sample paths of $B(t)$ in the mono-molecular chain model with $\tau = 2,4,8$ generated directly from a tau leaping method sample path with $\tau = 1$. The algorithm can be thought of as generating multiple approximations of the same exact biochemical reaction network sample path. The algorithm is the \emph{correlated tau leaping method}:
\begin{enumerate}
	\item initialise time $t  = 0$, and states $\bvec{Z}_\ell$, $\bvec{Z}_{\ell-1}$ corresponding to sample paths with $\tau = \tau_{\ell}$ and $\tau = \tau_{\ell-1}$, respectively;
	\item if $t + \tau_\ell > T$, then terminate simulation, otherwise, continue\label{line:loop};
	\item calculate propensities for path $\bvec{Z}_\ell$,  $a_1(\bvec{Z}_\ell), \ldots, a_M(\bvec{Z}_\ell)$;
	\item if $t/\tau_{\ell}$ is not an integer multiple of $\delta$, then go to step~\ref{line:virtchannels}, otherwise continue;
	\item calculate propensities for path $\bvec{Z}_{\ell-1}$,  $a_1(\bvec{Z}_{\ell-1}), \ldots, a_M(\bvec{Z}_{\ell-1})$, initialise intermediate state $\bar{\bvec{Z}} = \bvec{Z}_{\ell-1}$;
	\item  for each reaction $j = 1,\ldots, M$\label{line:virtchannels};
		  \begin{enumerate}
		  	\item calculate virtual propensities, $b_{j,1} = \min\{a_j(\bvec{Z}_\ell),a_j(\bvec{Z}_{\ell-1})\}$, $b_{j,2} = a_j(\bvec{Z}_\ell) - b_{j,1}$, and $b_{j,3} = a_j(\bvec{Z}_{\ell-1}) - b_{j,1}$; 
		  	\item generate virtual reaction event counts, $Y_{j,1} \sim \text{Po}(b_{j,1}\tau_{\ell})$, $Y_{j,2} \sim \text{Po}(b_{j,2}\tau_{\ell})$, and $Y_{j,3} \sim \text{Po}(b_{j,3}\tau_{\ell})$;
		  \end{enumerate}
	\item set, $\bvec{Z}_\ell = \bvec{Z}_\ell + (Y_{1,1} + Y_{1,2})\boldsymbol{\nu}_1 +  \cdots + (Y_{M,1} + Y_{M,2})\boldsymbol{\nu}_M $;
	\item set, $\bar{\bvec{Z}} = \bar{\bvec{Z}} + (Y_{1,1} + Y_{1,3})\boldsymbol{\nu}_1 +  \cdots + (Y_{M,1} + Y_{M,3})\boldsymbol{\nu}_M $;
	\item if $(t+ \tau_{\ell})/\tau_{\ell}$ is an integer multiple of $\delta$, then Set $\bvec{Z}_{\ell-1} = \bar{\bvec{Z}}$;
	\item update time $t = t + \tau_{\ell}$, and go to step~\ref{line:loop}.
\end{enumerate}
See example implementation, \texttt{CorTauLeapingMethod.m}, and example usage \texttt{DemoCorTauLeap.m}.

Given the correlated tau leaping method, Monte Carlo estimation can be applied to each term in~\eqref{eq:mlmc} to give
\begin{equation}
\hat{\bvec{Z}}_L(T) = \frac{1}{n_0}\sum_{i=1}^{n_0} \bvec{Z}_0(T)^{(i)} + \sum_{\ell=1}^{L}\frac{1}{n_\ell}\sum_{i=1}^{n_\ell}\left[\bvec{Z}_{\ell}(T)^{(i)}-\bvec{Z}_{\ell-1}(T)^{(i)}\right],
\end{equation}
where $\bvec{Z}_0(T)^{(1)}, \ldots, \bvec{Z}_0(T)^{(n_0)}$ are $n_0$ independent tau leaping method sample paths with $\tau = \tau_0$, and $(\bvec{Z}_{\ell}(T)^{(1)}$, $\bvec{Z}_{\ell-1}(T)^{(1)}), \ldots, (\bvec{Z}_{\ell}(T)^{(n_\ell)},\bvec{Z}_{\ell-1}(T)^{(n_\ell)})$ are $n_\ell$ paired correlated tau leaping method sample paths with time steps $\tau = \tau_{\ell}$, $\tau = \tau_{\ell-1}$ and $\tau_{\ell-1} = \delta \tau_{\ell}$ for each bias correction $\ell = 1,2, \ldots, L$. 
Given an error tolerance, $h$, it is possible to calculate an optimal sequence of sample path numbers $n_0,n_1,\ldots, n_L$ such that the total computation time is optimised~\cite{Anderson2012,Lester2016,Giles2008}. The results are shown in Figure~\ref{fig:MLMC}(a) for a more computationally challenging parameter set of the mono-molecular chain model. See the example implementation, \texttt{MultilevelMonteCarlo.m} and \texttt{DemoMonteCarlo.m}, for the full performance comparison.

As formulated here, MLMC results in a biased estimator, though it is significantly more efficient to reduce the bias of this estimator than by direct use of the tau leaping method. If an unbiased estimator is required, then this can be achieved by correlating exact SSA sample paths with approximate SSA sample paths. Anderson and Higham~\cite{Anderson2012} demonstrate a method for correlating tau leaping method sample paths and modified next reaction method sample paths, and Lester et al.~\cite{Lester2016} demonstrate correlating tau leaping method sample paths and Gillespie direct method sample paths. Further refinements such as adaptive and non-nested $\tau_{\ell}$ steps are also considered by Lester et al.~\cite{Lester2015}, a multilevel hybrid scheme is developed by Moraes et~al.~\cite{Moraes2016} and Wilson and Baker~\cite{Wilson2016} use MLMC and maximum entropy methods to generate approximations to the chemical master equation.

\subsection{Summary of the forwards problem}
Significant progress has be made in the study of computational methods for the solution to the forwards problem.  
As a result, forwards problem is relatively well understood, particularly for well-mixed systems, such as the biochemical reaction network models we consider in this review.
  
 An exact solution to simulation is achieved though the development of exact SSAs. However, if Monte Carlo methods are required to determine expected behaviours, then exact SSAs can be computationally burdensome. While approximate SSAs provide improvements is some cases, highly accurate estimates will still often become burdensome since very small time steps will be required to keep the bias at the same order as the estimator standard deviation. In this respect, MLMC methods provide impressive computational improvements without any loss in accuracy. Such methods have become popular in financial applications~\cite{Giles2015,Higham2015}, however, there have been fewer examples in a biological setting.

 Beyond the Gillespie direct method, the efficiency of sample path generation has been dramatically improved through the advancements in both exact SSAs and approximate SSAs. While approximate SSAs like the tau leaping method provide computational advantages, they also introduce approximations. Some have noted that the error in these approximations likely to be significantly lower than the modelling error compared with the real biological systems~\cite{Wilkinson2009}. However, there is no general theory or guidelines as to when approximate SSAs are safe to use for applications. 
 
 We have only dealt with stochastic models that are well-mixed, that is, spatially homogeneous. The development of robust theory and algorithms accounting for spatial heterogeneity is still an active area of research~\cite{Gillespie2013,Schnoerr2017,Smith2016}. The model of biochemical reaction networks, based on the chemical master equation, can be extended to include spatial dynamics through the \emph{reaction-diffusion master equation}~\cite{Isaacson2008}. However, care must be taken in its application because the kinetics of the reaction-diffusion master equation depend on the spatial discretisation and it not always guaranteed to converge in the continuum limit~\cite{Erban2009,Fange2010,Gillespie2014,Isaacson2008,Smith2016}. We refer the reader to Gillespie et al.~\cite{Gillespie2013} and Isaacson~\cite{Isaacson2008} for useful discussions on this topic.
 
 State-of-the-art Monte Carlo schemes, such as MLMC methods, have the potential to significantly accelerate the computation of summary statistics for the exploration of the range of biochemical reaction network behaviours.
  However, these approaches are also known to perform poorly for some biochemical reaction network models~\cite{Anderson2012}. An open question is related to the characterisation of biochemical reaction network models that will benefit from a MLMC scheme for summary statistic estimation. Furthermore, to our knowledge there has been no application of the MLMC approach to the spatially heterogeneous case. The potential performance gains make MLMC a promising space for future development.

\section{The inverse problem}
\label{sec:Inv}

When applying stochastic biochemical reaction network models to real applications, one often wants to perform statistical inference to estimate the model parameters. That is, given experimental data, and a biochemical reaction network model, the inverse problem seeks to infer the kinetic rate parameters and quantify the uncertainty in those estimates.
Just as with the forwards problem, an enormous volume of literature has been dedicated to the problem of inference in stochastic biochemical reaction network models. Therefore, we cannot cover all computational methods in detail. Rather we focus on a computational Bayesian perspective. For further reading, the monograph by Wilkinson~\cite{Wilkinson2012} contains very accessible discussions on inference techniques in a systems biology setting, also the monographs by Gelman et~al.~\cite{Gelman2014} and Sisson et~al.~\cite{Sisson2018} contain a wealth of information on Bayesian methods more generally.

\subsection{Experimental techniques}

Typically, time course data are derived from time-lapse microscopy images and fluorescent reporters~\cite{Finkenstadt2008,Iafolla2008,Wilkinson2009}. Advances in microscopy and fluorescent technologies are enabling intracellular processes to be inspected at unprecedented resolutions~\cite{Chen2014,Bajar2016,Leung2011,Sahl2017}. Despite these advances, the resulting data never provide complete observations since: (i) the number of chemical species that may be observed concurrently is relatively low~\cite{Wilkinson2009}; (ii) two chemical species might be indistinguishable from each other~\cite{Golightly2011}; and (iii) the relationships between fluorescence levels and actual chemical species copy numbers may not be direct, in particular, the degradation of a protein may be more rapid than that of the fluorescent reporter~\cite{Iafolla2008,Vittadello2018}. That is, inferential methods must be able to deal with uncertainty in observations. 

For the purposes of this review, we consider time-course data. Specifically,  we suppose the data consists of $n_t$ observations of the biochemical reaction network state vector at discrete points in time, $t_1, t_2, \ldots, t_{n_t}$. That is, $\bvec{Y}_{\text{obs}} = \left[\bvec{Y}(t_1),\bvec{Y}(t_2),\ldots,\bvec{Y}(t_{n_t})\right]$, where $\bvec{Y}(t)$ represents an observation of the state vector sample path $\bvec{X}(t)$. To model observational uncertainty, it is common to treat observations as subject to additive noise~\cite{Finkenstadt2008,Golightly2011,Schnoerr2017,Toni2009}, so that
\begin{equation}
\label{eq:obs}
\bvec{Y}(t) = \bvec{A} \bvec{X}(t) + \boldsymbol{\xi},
\end{equation} 
where $\bvec{A}$ is a $K \times N$ matrix and $\boldsymbol{\xi}$ is a $K \times 1$ vector of independent Gaussian random variables. The observation vectors, $\bvec{Y}(t)$, are $K \times 1$ vectors, with $K \leq N$, reflecting the fact that only a sub-set of chemical species of $\bvec{X}(t)$ are generally observed, or possibly only a linear combination of chemical species~\cite{Golightly2011}. The example code, \texttt{GenerateObservations.m}, simulates this observation process (\eqref{eq:obs}) given a fully specified biochemical reaction network model. 

\subsubsection{Example data}
The computation examples given in this review are based on two synthetically generated data sets, corresponding to the biochemical reaction network models given in Section~\ref{sec:BCRN}. This enables the comparison between inference methods and the accuracy of inference. 

The data for inference on the mono-molecular chain model (\eqref{eq:monomol}) are taken as perfect observations, that is, $K = N$, $\bvec{A} = \bvec{I}$ and $\Prob{\boldsymbol{\xi} = \bvec{0}} = 1$.  A single sample path is generated for the mono-molecular chain model with true parameters, $\paramvec_{\text{true}} = [1.0,0.1,0.05]$, and initial condition $A(0) = 100$ and $B(0) = 0$ using the Gillespie direct method. Observations are made using \eqref{eq:obs} applied at $n_t = 4$ discrete times, $t_1 = 25$, $t_2 = 50$, $t_3 = 75$ and $t_4 = 100$. The resulting data are given in Appendix~\ref{sec:app_data}. 

The data for inference on the enzyme kinetic model (\eqref{eq:michment}) assumes incomplete and noisy observations. Specifically, only the product is observed, so $K = 1$, $\bvec{A} = [0,0,0,1]$. Further, we assume that there is some measurement error, $\xi \sim \mathcal{N}(0,4)$; that is, the error standard deviation is two product molecules. The data is generated using the Gillespie direct method with true parameters $\paramvec_{\text{true}} = [0.001,0.005,0.01]$ and initial condition $E(0) = 100$, $S(0) = 100$, $C(0) = 0$ and $P(0) = 0$. \eqref{eq:obs} is evaluated at $n_t = 5$ discrete times,  $t_1 = 0$, $t_2 = 20$, $t_3 = 40$, $t_4 = 60$ and $t_5 = 80$ (including an observation of the initial state), yielding the data in Appendix~\ref{sec:app_data}.

\subsection{Bayesian inference}

 Bayesian methods have been demonstrated to provide a powerful framework for the design of experiments, model selection and parameter estimation, especially in the life sciences~\cite{Browning2017,Ellison2004,Liepe2013,Maclaren2015,Maclaren2017,Vanlier2014,Warne2019,Warne2017a}. Given observations,  $\bvec{Y}_{\text{obs}}$, and a biochemical reaction network model parameterised by the $M \times 1$  real-valued vector of kinetic parameters, $\paramvec = [k_1,k_2,\ldots,k_M]^T$, the task is to quantify knowledge of the true parameter values in light of the data and prior knowledge.  
This is expressed mathematically through Bayes' Theorem,
\begin{equation}
\label{eq:bayes}
\CondPDF{\paramvec}{\bvec{Y}_{\text{obs}}} = \frac{\CondPDF{\bvec{Y}_{\text{obs}}}{\paramvec}\PDF{\paramvec}}{\PDF{\bvec{Y}_{\text{obs}}}}.
\end{equation}
The terms in \eqref{eq:bayes} are interpreted as follows: $\CondPDF{\paramvec}{\bvec{Y}_{\text{obs}}}$ is the \emph{posterior} distribution, that is, the probability\footnote{Since $\paramvec$ and $\bvec{Y}_{\text{obs}}$ are real-valued, we are not really dealing with a probability distribution functions, but rather probability \emph{density} functions. We will not continue to make this distinction. However, the main technicality is that it not longer makes sense to say ``the probability of $\paramvec = \paramvec_0$'' but rather only ``the probability that $\paramvec$ is close to $\paramvec_0$''. The probability density function must be integrated over the region ``close to $\paramvec_0$'' to yield this probability.} of parameter combinations, $\paramvec$, given the data, $\bvec{Y}_{\text{obs}}$; $\PDF{\paramvec}$ is the \emph{prior} distribution, that is, the probability of parameter combinations before taking the data into account; $\CondPDF{\bvec{Y}_{\text{obs}}}{\paramvec}$ is the \emph{likelihood}, that is, the probability of observing the data given a parameter combination; and $\PDF{\bvec{Y}_{\text{obs}}}$ is the \emph{evidence}, that is, the probability of observing the data over all possible parameter combinations.
Assumptions about the parameters and the biochemical reaction network model are encoded through the prior and the likelihood, respectively. The evidence acts as a normalisation constant, and ensures the posterior is a true probability distribution\footnote{That is, the probability of an event that encompasses all possible outcomes is one.}. 

First, consider the special case $\bvec{Y}(t) = \bvec{X}(t)$, that is, the biochemical reaction network state can be perfectly observed at time $t$. In this case, the likelihood is
\begin{equation}
\label{eq:lh}
\CondPDF{\bvec{Y}_{\text{obs}}}{\paramvec} = \prod_{i=1}^{n_t} P\left(\bvec{Y}(t_i),t_i - t_{i-1} \mid \bvec{Y}(t_{i-1})\right),
\end{equation} 
where the function $P$ is the solution to the chemical master equation (\eqref{eq:cme}) and $t_0 = 0$~\cite{Browning2019,Wilkinson2011}. It should be noted that, due to the stochastic nature of $\bvec{X}(t)$, this perfect observation case is unlikely to recover the true parameters.  Regardless of this issue, since the likelihood depends on the solution to the chemical master equation, the exact Bayesian posterior will not be analytically tractable in any practical case. In fact, even for the mono-molecular chain model there are problems since the evidence term is not analytically tractable. The example code, \texttt{DemoDirectBayesCME.m}, provides an attempt at such a computation, though this code is not practical for an accurate evaluation. Just as with the forwards problem, we must defer to sampling methods and Monte Carlo.

\subsection{Sampling methods}
 For this review, we focus on the task of estimating the posterior mean,
\begin{equation}
\label{eq:postmean}
\CondE{\paramvec}{\bvec{Y}_{\text{obs}}} = \int_{\paramspace} \paramvec\, \CondPDF{\paramvec}{\bvec{Y}_{\text{obs}}}\, \text{d}\paramvec, 
\end{equation} 
where $\paramspace$ is the space of possible parameter combinations. However, the methods presented here are applicable to other quantities of interest. For example, the posterior covariance matrix,
\begin{eqnarray}
\label{eq:postcov}
\CondC{\paramvec}{\bvec{Y}_{\text{obs}}} = \int_{\paramspace} \left(\paramvec - \CondE{\paramvec}{\bvec{Y}_{\text{obs}}}\right)\left(\paramvec - \CondE{\paramvec}{\bvec{Y}_{\text{obs}}}\right)^T  \CondPDF{\paramvec}{\bvec{Y}_{\text{obs}}}\, \text{d}\paramvec,
\end{eqnarray}
is of interest as it provides an indicator of uncertainty associated with the inferred parameters. Marginal distributions are extremely useful for visualisation: the marginal posterior distribution of the $j$th kinetic parameter is 
\begin{equation}
\label{eq:margpost}
\CondPDF{k_j}{\bvec{Y}_{\text{obs}}} = \int_{\paramspace_{j}} \CondPDF{\paramvec}{\bvec{Y}_{\text{obs}}}\, \prod_{i \neq j}\text{d}k_i, 
\end{equation}
where $\paramspace_{j} \subset \paramspace$ is the parameter space excluding the $j$th dimension. 

 Just as with Monte Carlo for the forwards problem, we can estimate posterior expectations (shown here for \eqref{eq:postmean}, but the method may be similarly applied to \eqref{eq:postcov} and \eqref{eq:margpost}) using Monte Carlo,
\begin{equation}
\label{eq:postmc}
\CondE{\paramvec}{\bvec{Y}_{\text{obs}}} \approx \hat{\paramvec} = \frac{1}{m} \sum_{i=1}^m \paramvec^{(i)},
\end{equation}
where $\paramvec^{(1)},\ldots, \paramvec^{(m)}$ are independent samples from the posterior distribution. In the remainder of this section, we focus on computational schemes for generating such samples. We assume throughout that it is possible to generate samples from the prior distribution. 

It is important to note that the sampling algorithms we present are not directly relevant to statistical estimators that are not based on expectations, such as \textit{maximum likelihood estimators} or the \textit{maximum a posteriori}. However, these samplers can be modified through the use of \textit{data cloning}~\cite{Lele2007,Picchini2017} to approximate these effectively. Estimator variance and confidence intervals may also be estimated using bootstrap methods~\cite{Efron1979,Rubin1981}.    

\subsubsection{Exact Bayesian sampling}
Assuming the likelihood can be evaluated, that is, the chemical master equation is tractable, a na\"{i}ve method of generating $m$ samples from the posterior is the \emph{exact rejection sampler}:
\begin{enumerate}
	\item initialise index $i = 0$,
	\item generate a prior sample $\paramvec^* \sim \PDF{\paramvec}$ \label{line:sampleprior};
	\item calculate acceptance probability $\alpha = \CondPDF{\bvec{Y}_{\text{obs}}}{\paramvec^*}$;
	\item with probability $\alpha$, accept $\paramvec^{(i+1)} = \paramvec^*$ and $i = i+1$;
	\item if $i = m$, terminate sampling, otherwise go to step~\ref{line:sampleprior};
\end{enumerate}
Unsurprisingly, this approach is almost never viable as the likelihood probabilities are often extremely small. In the code example, \texttt{DemoExactBayesRejection.m}, the acceptance probability is never more than $9\times 10^{-15}$. 

The most common solution is to construct a type of stochastic process (a Markov chain) in the parameter space to generate $m_n$ steps, $\paramvec^{(0)},\ldots, \paramvec^{(m_n)}$. An essential property of the Markov chain used is that its stationary distribution is the target posterior distribution. This approach is called Markov chain Monte Carlo (MCMC), and a common method is the \emph{Metropolis-Hastings method}~\cite{Metroplis1953,Hastings1970}:
\begin{enumerate}
	\item initialise $n = 0$ and select starting point $\paramvec^{(0)}$;
	\item generate a proposal sample, $\paramvec^{*} \sim \Kernel{\paramvec}{\paramvec^{(n)}}$\label{line:sampleprop};
	\item calculate acceptance probability $\alpha = \min\left(1, \dfrac{\CondPDF{\bvec{Y}_{\text{obs}}}{\paramvec^{*}}\PDF{\paramvec^{*}}\Kernel{\paramvec^{(n)}}{\paramvec^{*}}}{\CondPDF{\bvec{Y}_{\text{obs}}}{\paramvec^{(n)}}\PDF{\paramvec^{(n)}}\Kernel{\paramvec^{*}}{\paramvec^{(n)}}}\right)$\label{line:MHaccept};
	\item with probability $\alpha$, set $\paramvec^{(n+1)} = \paramvec^{*}$, and with probability $1-\alpha$, set $\paramvec^{(n+1)} = \paramvec^{(n)}$;
	\item update time, $n = n + 1$;
	\item if $n > m_n$, terminate simulation, otherwise go to step~\ref{line:sampleprop}; 	 
\end{enumerate}
 The acceptance probability is based on the relative likelihood between two parameter configurations, the current configuration $\paramvec^{(n)}$ and a proposed new configuration $\paramvec^{*}$, as generated by the user-defined proposal kernel $\Kernel{\paramvec}{\paramvec^{(n)}}$. 
 It is essential to understand that since the samples, $\paramvec^{(0)},\ldots, \paramvec^{(m_n)}$, are produced by a Markov chain we cannot treat the $m_n$ steps as $m_n$ independent posterior samples. Rather, we need to take $m_n$ to be large enough that the $m_n$ steps are effectively equivalent to $m$ independent samples.  
 
 One challenge in MCMC sampling is the selection of a proposal kernel such that the size of $m_n$ required for the Markov chain to reach the stationary distribution, called the \textit{mixing time}, is small~\cite{Geyer1992}. If the variance of the proposal is too low, then the acceptance rate is high, but the mixing is poor since only small steps are ever taken. On the other hand, a proposal variance that is too high will almost always make proposals that are rejected, resulting in many wasted likelihood evaluations before a successful move event. Selecting good proposal kernels is a non-trivial exercise and we refer the reader to Cotter et~al.~\cite{Cotter2013}, Green et~al.~\cite{Green2015}, and Roberts and Rosenthal~\cite{Roberts2009} for detailed discussions on the wide variety of MCMC samplers including proposal techniques. Other techniques used to reduce correlations in MCMC samples include discarding the first $m_b$ steps, called \textit{burn-in} iterations, or sub-sampling the chain by using only every $m_h$th step, also called \textit{thinning}. However, in general, the use of thinning decreases the statistical efficiency of the MCMC estimator~\cite{Link2011,MacEachern1994}.

Alternative approaches for exact Bayesian sampling can also be based on \textit{importance sampling}~\cite{Glynn1989,Wilkinson2012}. Consider a random variable that cannot be simulated, $X \sim p(x)$, but suppose that it is possible to simulate another random variable, $Y \sim q(y)$. If $X,Y \in \Omega$ and $p(x) = 0$ whenever $q(y) = 0$, then
\begin{equation}
\E{X} = \int_\Omega x p(x)\,\text{d}x = \int_\Omega x \frac{p(x)}{q(x)} q(x)\,\text{d}x \approx \frac{1}{m} \sum_{i=1}^m \frac{p(Y^{(i)})}{q(Y^{(i)})}Y^{(i)},
\label{eq:is}
\end{equation}
where $Y^{(1)},\ldots, Y^{(m)}$ are independent samples of $q(Y)$. Using \eqref{eq:is}, one can show that if the distributions of the target, $p(x)$, and the proposal, $q(y)$, are similar, then a collection of samples, $Y^{(1)},\ldots, Y^{(m)}$, can be used to generate $m$ approximate samples from $p(x)$. This is called \emph{importance resampling}: 
\begin{enumerate}
	\item generate samples $Y^{(1)},\ldots, Y^{(m)} \sim q(y)$;
	\item compute weights $w^{(1)} = p(Y^{(1)})/q(Y^{(1)}),\ldots, w^{(m)} = p(Y^{(m)})/q(Y^{(m)})$; for $i=1,2,\ldots,m$;
	\item generate samples $\{X^{(1)},\ldots, X^{(m)}\}$ by drawing from $\{Y^{(1)},\ldots, Y^{(m)}\}$ with replacement using probabilities $\Prob{X = Y^{(i)}} = w^{(i)}\left/ \sum_{i=1}^m w^{(i)}\right.$.
\end{enumerate}
In Bayesian applications the prior is often very different to the posterior. In such a case, importance resampling may be applied using a sequence of intermediate distributions. This \textit{sequential importance resampling} and is one approach from the family of \textit{sequential Monte Carlo} (SMC)~\cite{DelMoral2006} samplers. However, like the Metropolis-Hastings method, these methods also require explicit calculations of the likelihood function in order to compute the weights, thus all of these approaches are infeasible for practical biochemical reaction networks. Therefore, we only present more practical forms of these methods later.

More recently, it has been shown that the MLMC telescoping summation can accelerate the computation of posterior expectations when using MCMC~\cite{Dodwell2015,Efendiev2015} or SMC~\cite{Beskos2016}. The key challenges in these applications is the development of appropriate coupling strategies. We do not cover these technical details in this review.

\subsubsection{Likelihood-free methods}
Since exact Bayesian sampling is rarely tractable, due to the intractability of the chemical master equation, alternative, likelihood-free sampling methods are required. Two main classes of likelihood-free methods exist: (i) so-called pseudo-marginal MCMC; and (ii) approximate Bayesian computation (ABC).
The focus of this review is ABC methods, though we first briefly discuss the pseudo-marginal approach. 

The basis of the pseudo-marginal approach is to use a MCMC sampler (e.g., Metropolis-Hastings method), but replace the explicit likelihood evaluation with a likelihood estimate obtained through Monte Carlo simulation of the forwards problem~\cite{Andrieu2009}. For example, a direct unbiased Monte Carlo estimator is  
\begin{align*}
\CondPDF{\bvec{Y}_{\text{obs}}}{\paramvec} &= \int_{\Omega^{n_t}}\prod_{i=1}^{n_t} \CondPDF{\bvec{Y}(t_i)}{\bvec{X}(t_i)}P\left(\bvec{X}(t_i),t_i - t_{i-1} \mid \bvec{X}(t_{i-1})\right)\, \text{d}\bvec{X}(t_i)\\
&\approx \frac{1}{n}\sum_{j=1}^n \prod_{i=1}^{n_t} \CondPDF{\bvec{Y}(t_i)}{\bvec{X}(t_i)^{(j)}},
\end{align*}
where $[\bvec{X}(t_1)^{(j)},\ldots,\bvec{X}(t_{n_t})^{(j)}]^{T}$ for $j= 1,2, \ldots, n$ are independent sample paths of the biochemical reaction network of interest observed discretely at times $t_1, t_2,\ldots, t_{n_t}$. The most successful class of pseudo-marginal techniques, particle MCMC~\cite{Andrieu2010}, apply a SMC sampler to approximate the likelihood and inform the MCMC proposal mechanism. The particle marginal Metropolis-Hastings method is a popular variant~\cite{Andrieu2010,Golightly2011}. However, the recent model based proposals variant also holds promise for biochemical reaction networks specifically~\cite{Pooley2015}. 

A particularly nice feature of pseudo-marginal methods is that they are unbiased\footnote{Provided the method of estimating the likelihood is unbiased, or has bias that is independent of $\paramvec$.}, that is, the Markov chain will still converge to the exact target posterior distribution. This property is sometimes called an ``exact approximation''~\cite{Golightly2011,Wilkinson2012}. Unfortunately, the Markov chains in these methods typically converge more slowly than their exact counterparts. However, computational improvements have been obtained through application of MLMC~\cite{Jasra2018}.

Another popular likelihood-free approach is ABC~\cite{Pritchard1999,Sisson2018,Tavare1997}. ABC methods have enabled very complex models to be investigated, that are otherwise intractable~\cite{Sunnaker2013}. Furthermore, ABC methods are very intuitive, leading to wide adoption within the scientific community~\cite{Sunnaker2013}. Applications are particularly prevalent in the life sciences, especially in evolutionary biology~\cite{Beaumont2002,Pritchard1999,Ratmann2012,Tavare1997,WilkinsonR2011}, cell biology~\cite{Johnston2014,Ross2017,Vo2015}, epidemiology~\cite{Tanaka2006}, ecology~\cite{Csillery2010,Stumpf2014}, and systems biology~\cite{Toni2009,Wilkinson2011}.

The basis of ABC is a discrepancy metric $\discrep{\bvec{Y}_{\text{obs}}}{\bvec{S}_{\text{obs}}}$ that provides a measure of closeness between the data, $\bvec{Y}_{\text{obs}}$, and simulated data $\bvec{S}_{\text{obs}}$ generated through stochastic simulation of the biochemical reaction network with simulated measurement error. Thus, acceptance probabilities are replaced with a deterministic acceptance condition, $\discrep{\bvec{Y}_{\text{obs}}}{\bvec{S}_{\text{obs}}} \leq \epsilon$, where $\epsilon$ is the discrepancy threshold. This yields an approximation to Bayes' Theorem,
\begin{align}
\label{eq:approxbayes}
\CondPDF{\paramvec}{\bvec{Y}_{\text{obs}}} &\approx \CondPDF{\paramvec}{\discrep{\bvec{Y}_{\text{obs}}}{\bvec{S}_{\text{obs}}} \leq \epsilon} \notag\\
&= \frac{\CondPDF{\discrep{\bvec{Y}_{\text{obs}}}{\bvec{S}_{\text{obs}}} \leq \epsilon}{\paramvec}\PDF{\paramvec}}{\PDF{\discrep{\bvec{Y}_{\text{obs}}}{\bvec{S}_{\text{obs}}} \leq \epsilon}}.
\end{align}
The key insight here is that the ability to produce sample paths of a biochemical reaction network model, that is, the forwards problem, enables an approximate algorithm for inference, that is, the inverse problem, regardless of the intractability or complexity of the likelihood. In fact, a formula for the likelihood need not even be known.    

\begin{figure}[h]
	\centering
	\includegraphics[width=\linewidth]{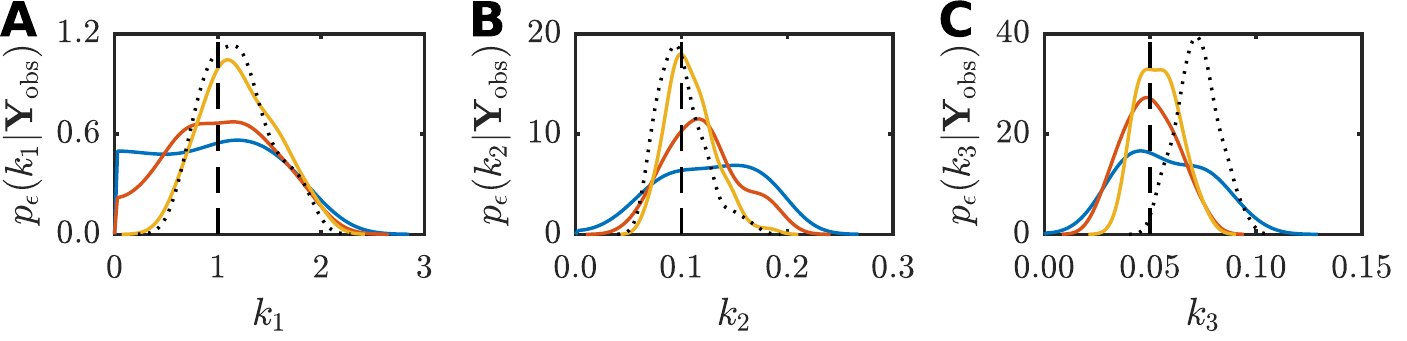}
	\caption{Convergence of ABC posterior to the true posterior as $\epsilon \to 0$ for the mono-molecular chain inference problem. Marginal posteriors are plotted for $\epsilon = 50$ (blue solid), $\epsilon = 25$ (red solid), $\epsilon = 12.5$ (yellow solid), and $\epsilon = 0$ (black dotted). Here, the $\epsilon = 0$ case corresponds to the exact likelihood using the chemical master equation solution. (a) marginal posteriors for $k_1$, (b) marginal posteriors for $k_2$, and (c) marginal posteriors for $k_3$. The true parameter values (black dashed) are $k_1 = 1.0$, $k_2 = 0.1$ and $k_3 = 0.05$. Note that the exact Bayesian posterior does not recover the true parameter  for $k_3$. The priors used are $k_1 \sim \mathcal{U}(0,3)$,$k_2 \sim \mathcal{U}(0,0.3)$, and $k_3 \sim \mathcal{U}(0,0.15)$.}
	\label{fig:ABCconv}
\end{figure}

The discrepancy threshold determines the level of approximation, however, under the assumption of model and observation error, \eqref{eq:approxbayes} can be treated as exact~\cite{Wilkinson2013}. As $\epsilon \to 0$ then $\CondPDF{\paramvec}{\discrep{\bvec{Y}_{\text{obs}}}{\bvec{S}_{\text{obs}}} \leq \epsilon} \to \CondPDF{\paramvec}{\bvec{Y}_{\text{obs}}}$~\cite{Barber2015,Fearnhead2012}. Using data for the mono-molecular chain model, we can demonstrate this convergence, as shown in Figure~\ref{fig:ABCconv} (see also Appendix~\ref{sec:app_results}). The marginal posterior distributions are shown for each parameter, for various values of $\epsilon$ and compared with the exact marginal posteriors. 
The discrepancy metric used is
\begin{equation}
\label{eq:discrep}
\discrep{\bvec{Y}_{\text{obs}}}{\bvec{S}_{\text{obs}}} = \left[\sum_{i=1}^{n_t} (\bvec{Y}(t_i) - \bvec{S}(t_i))^2\right]^{1/2},
\end{equation} 
where $\bvec{S}(t)$ is simulated data generated using Gillespie direct method and \eqref{eq:obs}. The example code, \texttt{DemoABCConvergence.m}, is used to generate these marginal distributions. 

For any $\epsilon > 0$, ABC methods are biased, just as the tau leaping method is biased for the forwards problem. Therefore, a Monte Carlo estimate of a posterior summary statistic, such as \eqref{eq:postmc}, needs to take this into bias account. Just like the tau leaping method, the rate of convergence in \emph{mean-square} of ABC based Monte Carlo is degraded because of this bias~\cite{Barber2015,Fearnhead2012}. Furthermore, as the dimensionality of the data increases, small values of $\epsilon$ are not computationally feasible.  In such cases, the data dimensionality may be reduced by computing lower dimensional summary statistics~\cite{Sunnaker2013}, however, these must be \textit{sufficient statistics} in order to ensure the ABC posterior still converges to the correct posterior as $\epsilon \to 0$. We refer the reader to Fearnhead and Prangle~\cite{Fearnhead2012} for more detail on this topic.

\subsubsection{Samplers for approximate Bayesian computation}

We now focus on computational methods for generating $m$ samples, $\paramvec_\epsilon^{(1)}, \ldots, \paramvec_\epsilon^{(m)}$, from the ABC posterior (\eqref{eq:approxbayes}) with $\epsilon > 0$ and discrepancy metric as given in \eqref{eq:discrep}. Throughout, we denote $s(\bvec{S}_{\text{obs}}; \paramvec)$, as the process for generating simulated data given a parameter vector, this process is identical to the processes used to generate our synthetic example data. 

In general, the ABC samplers are only computationally viable if the data simulation process is not computationally expensive, in the sense that it is feasible to generate millions of sample paths. However, this is not always realistic, and many extensions exist to standard ABC in an attempt to deal with these more challenging cases. The \textit{Lazy ABC} method~\cite{Prangle2014} applies an additional probability rule that terminates simulations early if it is likely that $\discrep{\bvec{Y}_{\text{obs}}}{\bvec{S}_{\text{obs}}} > \epsilon$. The \textit{approximate ABC} method~\cite{Buzbas2015,Lambert2018} utilises a small set of data simulations to construct an approximation to the data simulation process.

\begin{landscape}
	\begin{figure}[h]
		\centering
		\includegraphics[width=\linewidth]{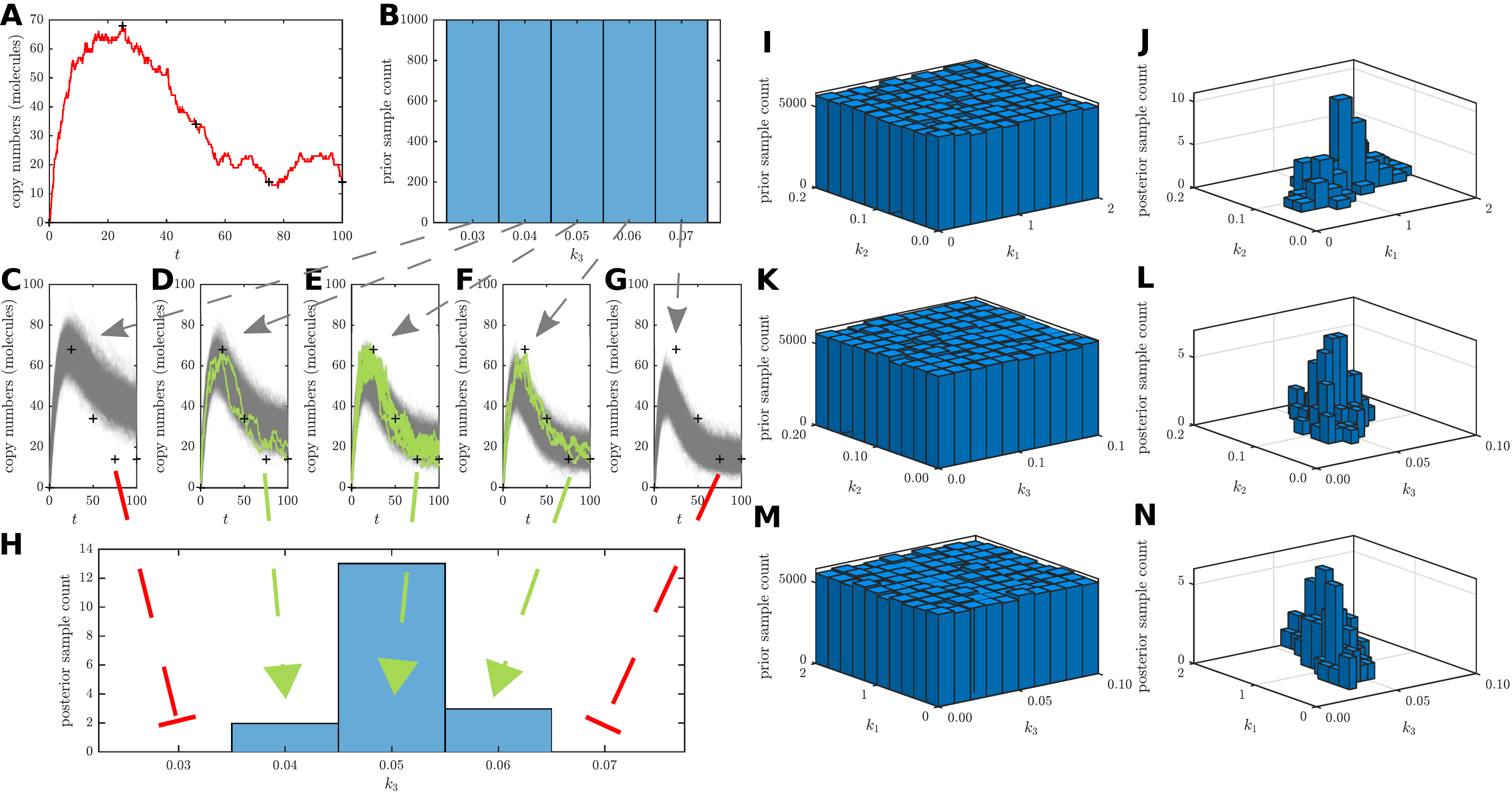}
		\caption{Demonstration of the ABC rejection sampler method using the mono-molecular chain model data set. (a) Experimental observations (black crosses) obtained from a true sample path of $B(t)$ in the mono-molecular chain model (red line) with $k_1 = 1.0$, $k_2 = 0.1$ and $k_3 = 0.05$, and initial conditions, $A(0) = 100$ and $B(0) = 0$. (b) Prior for inference on $k_3$. (c)--(g) Stochastic simulation of many choices of $k_3$ drawn from the prior, showing accepted (solid green) and rejected (solid gray) sample paths with $\epsilon = 15$ (molecules). (h) ABC posterior for $k_3$ generated from accepted samples. (i)--(n) Bivariate marginal distributions of the full ABC inference problem on $\paramvec = \left\{k_1,k_2,k_3\right\}$.}
		\label{fig:ABCproc}
	\end{figure}
\end{landscape}

The most notable early applications of ABC samplers are Beaumont et al.~\cite{Beaumont2002}, Pritchard et~al.~\cite{Pritchard1999}, and Taver\'{e} et al.~\cite{Tavare1997}. The essential development of this early work is that of the \emph{ABC rejection sampler}:
\begin{enumerate}
	\item initialise index $i = 0$;
	\item generate a prior sample $\paramvec^* \sim \PDF{\paramvec}$ \label{line:samplepriorABC};
	\item generate simulated data, $\bvec{S}_{\text{obs}}^* \sim s(\bvec{S}_{\text{obs}}; \paramvec^*)$;
	\item if $\discrep{\bvec{Y}_{\text{obs}}}{\bvec{S}_{\text{obs}}^*} \leq \epsilon$, accept $\paramvec_\epsilon^{(i+1)} = \paramvec^*$ and set $i = i+1$, otherwise, continue;
	\item if $i = m$, terminate sampling, otherwise go to step~\ref{line:samplepriorABC};
\end{enumerate}
There is a clear connection with exact rejection sampler. Note that every accepted sample of the ABC posterior corresponds to at least one simulation of the forwards problem as shown in Figure~\ref{fig:ABCproc}. As a result, the computational burden of the inverse problem is significantly higher than the forwards problem, especially for small $\epsilon$. The example code, \texttt{ABCRejectionSampler.m}, provides an implementation of the ABC rejection sampler. 

Unfortunately, for small $\epsilon$, the computational burden of the ABC rejection sampler may be prohibitive as the acceptance rate is very low (this is especially an issue for biochemical reaction networks with highly variable dynamics). For the example in Figure~\ref{fig:ABCconv}, $m=100$ posterior samples takes approximately one minute for $\epsilon = 25$, but nearly ten hours for $\epsilon = 12.5$. However, for $\epsilon = 12.5$ the marginal ABC posterior for $k_3$ has not yet converged.

Marjoram et~al.~\cite{Marjoram2003} provide a solution via an ABC modification to Metropolis-Hastings method. This results in a Markov chain with the ABC posterior as the stationary distribution. This method is called \emph{ABC Markov chain Monte Carlo} (ABCMCMC):
\begin{enumerate} 
	\item initialise $n = 0$ and select starting point $\paramvec_\epsilon^{(0)}$;
	\item generate a proposal sample, $\paramvec^{*} \sim \Kernel{\paramvec}{\paramvec_\epsilon^{(n)}}$\label{line:samplepropABC};
	\item generate simulated data, $\bvec{S}_{\text{obs}}^* \sim s(\bvec{S}_{\text{obs}}; \paramvec^*)$;
	\item if $\discrep{\bvec{Y}_{\text{obs}}}{\bvec{S}_{\text{obs}}^*} > \epsilon$, then set $\paramvec_\epsilon^{(n+1)} = \paramvec_\epsilon^{(n)} $and go to step~\ref{line:loopABCMCMC}, otherwise continue\label{line:epscheck};
	\item calculate acceptance probability $\alpha = \min\left(1, \dfrac{\PDF{\paramvec^{*}}\Kernel{\paramvec_\epsilon^{(n)}}{\paramvec^{*}}}{\PDF{\paramvec_\epsilon^{(n)}}\Kernel{\paramvec^{*}}{\paramvec_\epsilon^{(n)}}}\right)$;
	\item with probability $\alpha$, set $\paramvec_\epsilon^{(n+1)} = \paramvec^{*}$, and with probability $1-\alpha$, set $\paramvec_\epsilon^{(n+1)} = \paramvec_\epsilon^{(n)}$;
	\item update time, $n = n + 1$\label{line:loopABCMCMC};
	\item if $n > m_n$, terminate simulation, otherwise go to step~\ref{line:samplepropABC}; 	 
\end{enumerate}
An example implementation, \texttt{ABCMCMCSampler.m}, is provided.

\begin{figure}
	\centering
	\includegraphics[width=\linewidth]{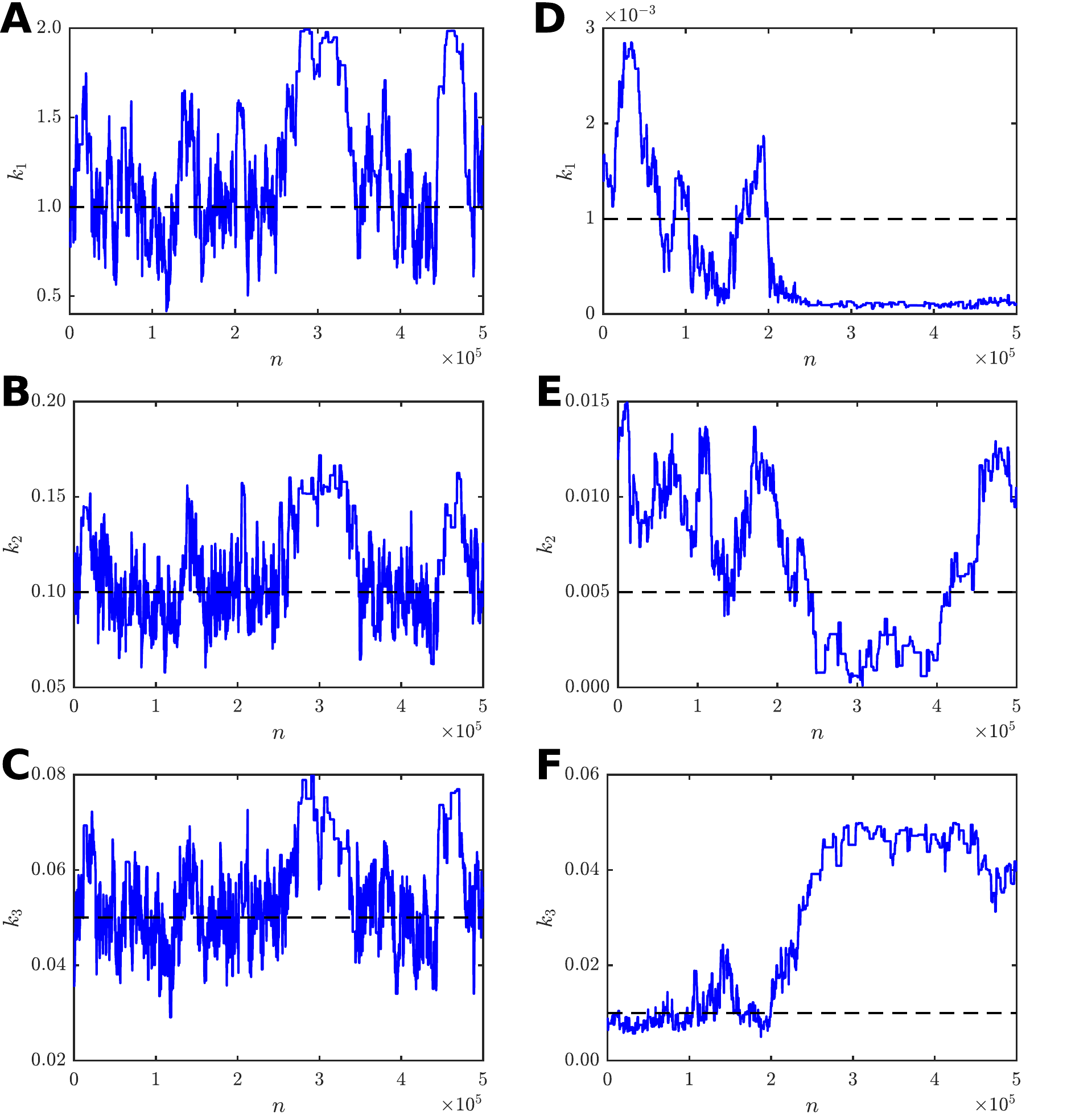}
	\caption{ $m_n = 500,000$ steps of ABCMCMC for: (a)--(c) the mono-molecular chain model; and (d)--(f) the enzyme kinetics model. True parameter values (dash black) are shown.}
	\label{fig:ABCMCMCconv}
\end{figure}

Just as with the Metropolis-Hastings method, the efficacy of the ABCMCMC rests upon the non-trivial choice of the proposal kernel, $\Kernel{\paramvec}{\paramvec_\epsilon^{(n)}}$. The challenge of constructing effective proposal kernels is equally non-trivial for ABCMCMC as for Metropolis-Hastings method. Figure~\ref{fig:ABCMCMCconv} highlights different Markov chain behaviours for heuristically chosen proposal kernels based on Gaussian random walks with variances that we alter until the Markov chain seems to be mixing well.

For the mono-molecular chain model (Figure~\ref{fig:ABCMCMCconv}(a)--(c)), ABCMCMC seems to be performing well with the Markov chain state moving largely in the regions of high posterior probability. However, for the enzyme kinetics model (Figure~\ref{fig:ABCMCMCconv}(d)--(f)), the Markov chain has taken a long excursion into a region of low posterior probability (see Figure~\ref{fig:ABCMCMCconv}(c)) where it is a long way from the true parameter value of $k_3 = 0.01$ for many steps. Such an extended path into the low probability region results in significant bias in this region and many many more steps of the algorithm are required to compensate for this~\cite{Sisson2007,Beaumont2009}. Just as with exact MCMC, burn-in samples and thinning may be used to reduce correlations but may not improve the final Monte Carlo estimate.

\newpage
To avoid the difficulties in ensuring efficient ABCMCMC convergence, Sisson et~al.~\cite{Sisson2007} developed an ABC variant of SMC. Alternative versions of the method are also deigned by Beaumont et al.~\cite{Beaumont2009} and Toni et al.~\cite{Toni2009}. The fundamental idea is to use sequential importance resampling to propagate $m_p$ samples, called \textit{particles}, through a sequence of $R+1$ ABC posterior distributions defined through a sequence of discrepancy thresholds, $\epsilon_0, \epsilon_1, \ldots,\epsilon_{R}$ with $\epsilon_r > \epsilon_{r+1}$ for $r = 0,1,\ldots, R-1$ and $\epsilon_0 = \infty$ (that is, $\CondPDF{\paramvec_{\epsilon_0}}{\bvec{Y}_{\text{obs}}}$ is the prior). This results in \emph{ABC sequential Monte Carlo} (ABCSMC):
\begin{enumerate}
	\item initialise $r = 0$ and weights $w_r^{(i)} = 1/m_p$ for $i = 1, \ldots, m_p$;
	\item generate $m_p$ particles from the prior, $\paramvec_{\epsilon_r}^{(i)} \sim \PDF{\paramvec}$, for $i = 1,2,\ldots,m_p$;
	\item set index $i = 0$ \label{line:SMCloop};
	\item randomly select integer, $j$, from the set $\{1,\ldots, m_p \}$ with \\$\Prob{j = 1} = w_{r}^{(1)}, \ldots, \Prob{j = m_p} = w_{r}^{(m_p)}$ \label{line:sampleweightSMC};
	\item generate proposal, $\paramvec^* \sim \Kernel{\paramvec}{\paramvec_{\epsilon_r}^{(j)}}$;
	\item generate simulated data, $\bvec{S}_{\text{obs}}^* \sim s(\bvec{S}_{\text{obs}}; \paramvec^*)$;
	\item if $\discrep{\bvec{Y}_{\text{obs}}}{\bvec{S}_{\text{obs}}^*} > \epsilon_{r+1}$,  go to step~\ref{line:sampleweightSMC}, otherwise continue; 
	\item set $\paramvec_{\epsilon_{r+1}}^{(i+1)*} = \paramvec^*$, $w_{r+1}^{(i+1)*} = \PDF{\paramvec^*}\left/\left[\sum_{j=1}^{m_p} w_{r}^{(j)}\Kernel{\paramvec^*}{\paramvec_{\epsilon_r}^{(j)}} \right]\right.$ and $i = i+1$;
	\item if $i < m_p$, go to step~\ref{line:sampleweightSMC}, otherwise continue;
	\item set index $i = 0$;
	\item randomly select integer, $j$, from the set $\{1,\ldots, m_p \}$ with \\ $\Prob{j = 1} = w_{r+1}^{(1)*}, \ldots, \Prob{j = m_p} = w_{r+1}^{(m_p)*}$ \label{line:resampleweightSMC};
	\item set $\paramvec_{r+1}^{i+1} = \paramvec_{r+1}^{(j)*}$ and $i = i+1$;
	\item if $i < m_p$, go to step~\ref{line:resampleweightSMC}, otherwise continue;
	\item set $w_{r+1}^{(i)} = 1/ \left[\sum_{j=1}^{m_p} w_{r+1}^{(j)*}\right]$ and $r = r +1$;
	\item if $r < R$, go to step~\ref{line:SMCloop}, otherwise terminate simulation;
\end{enumerate}
An implementation of the ABCSMC method is provided in \texttt{ABCSMCSampler.m}.

The ABCSMC method avoids some issues inherent in ABCMCMC such as strong sample correlation and long excursions into low probability regions. However, it is still the case that the number of particles, $m_p$, often needs to be larger than the desired number of independent samples, $m$, from the ABC posterior with $\epsilon_R$. This is especially true when there is a larger discrepancy between two successive ABC posteriors in the sequence. Although techniques exist to adaptively generate the sequence of acceptance thresholds~\cite{Drovandi2011}. Also note that ABCSMC still requires a proposal kernel to mutate the particles at each step and the efficiency of the method is affected by the choice of proposal kernel. Just as with ABCMCMC, the task of selecting an optimal proposal kernel is non-trivial~\cite{Filippi2013}.

The formulation of ABCSMC using a sequence of discrepancy thresholds hints that MLMC ideas could also be applicable.  Recently, a variety of MLMC methods for ABC have been proposed~\cite{Guha2017,Jasra2017,Warne2018}. All of these approaches are similar in their application of the multilevel telescoping summation to compute expectations with respect to ABC posterior distributions,
\begin{equation}
\label{eq:ABCMLMC}
\CondE{\paramvec_{\epsilon_{L}}}{\bvec{Y}_{\text{obs}}}  = \CondE{\paramvec_{\epsilon_{0}}}{\bvec{Y}_{\text{obs}}}  + \sum_{\ell=1}^{L}\CondE{\paramvec_{\epsilon_{\ell}} - \paramvec_{\epsilon_{\ell-1}} }{\bvec{Y}_{\text{obs}}}, 
\end{equation}
where $\epsilon_{0},\epsilon_{1},\ldots,\epsilon_{L}$ is a sequence of discrepancy thresholds with $\epsilon_{\ell} > \epsilon_{\ell+1}$ for all $\ell = 0,1,\ldots,L-1$.

Unlike the forwards problem, no exact solution has been found for the generation of correlated ABC posterior pairs, $(\paramvec_{\epsilon_\ell},\paramvec_{\epsilon_{\ell-1}})$, with $\ell = 1, \ldots,L$. Several approaches to this problem have been proposed: Guha and Tan~\cite{Guha2017} utilise a sequence of correlated ABCMCMC samplers in a similar manner to Dodwell et~al.~\cite{Dodwell2015} and Efendiev et~al.~\cite{Efendiev2015}; Jasra~et al.~\cite{Jasra2017} apply the MLSMC estimator of Beskos et~al.~\cite{Beskos2016} directly to \eqref{eq:ABCMLMC}; and Warne~et al.~\cite{Warne2018} develop a MLMC variant of ABC rejection sampler through the introduction of an approximation based on the empirical marginal posterior distributions. 

The coupling scheme of Warne et~al.~\cite{Warne2018} is the simplest to present succinctly. The scheme is based on the inverse transform method for sampling univariate distributions. Given random variable $X \sim \PDF{x}$, the cumulative distribution function is defined as,
\begin{equation*}
F(z) = \Prob{X \leq z} = \int_{-\infty}^z \PDF{x}\, \text{d}x.
\end{equation*}
Note that $0 \leq F(z) \leq 1$. If $F(z)$ has an inverse function, $F^{-1}(u)$, and $U^{(1)},U^{(2)},\ldots,U^{(m)}$ are independent samples of $\mathcal{U}(0,1)$, then independent samples from $p(x)$ can be generated using $X^{(1)} = F^{-1}(U^{(1)}), X^{(2)} = F^{-1}(U^{(2)}), \ldots, X^{(m)} = F^{-1}(U^{(m)})$.
 
Given samples, $\paramvec_{\epsilon_{\ell}}^{1}, \paramvec_{\epsilon_{\ell}}^{2}, \ldots, \paramvec_{\epsilon_{\ell}}^{m_\ell}$, an estimate for the posterior marginal cumulative distribution functions can be obtained using 
\begin{equation*}
F_{\ell,j}(z) \approx \hat{F}_{\ell,j}(z) = \frac{1}{m_\ell} \sum_{i=1}^{m_\ell} \ind{z}{k_{\epsilon_\ell,j}^{(i)}},
\end{equation*}
where $k_{\epsilon_\ell,j}^{(i)}$ is the $j$th component of $\paramvec_{\epsilon_\ell}^{(i)}$ for all $j = 1,\ldots, M$, and $\ind{z}{x} = 1$ if $x \leq z$ and $\ind{z}{x} = 0$ otherwise. Assuming we already have an estimate of the posterior marginal  cumulative distribution functions for acceptance threshold $\epsilon_{\ell-1}$, then we apply the transform $k_{\epsilon_{\ell-1},j}^{(i)} = \hat{F}^{-1}_{\ell-1,j}(\hat{F}_{\ell,j}(k_{\epsilon_{\ell},j}^{(i)}))$. This results in a coupled marginal pair $(k_{\epsilon_{\ell},j},k_{\epsilon_{\ell-1},j})^{(i)}$ since it is equivalent to using the inverse transform method to generate $k_{\epsilon_{\ell},j}^{(i)}$ and $k_{\epsilon_{\ell-1},j}^{(i)}$ using the same uniform sample $U^{(i)} \sim \mathcal{U}(0,1)$. This result leads to the \emph{approximately correlated ABC rejection sampler}:
\begin{enumerate}
	\item generate sample from level $\ell$  ABC posterior $\paramvec_{\epsilon_{\ell}} \sim \CondPDF{\paramvec_{\epsilon_{\ell}}}{\bvec{Y}_\text{obs}}$ using the ABC rejection sampler;
	\item set $k_{\epsilon_{\ell-1},j} = \hat{F}^{-1}_{\ell-1,j}(\hat{F}_{\ell,j}(k_{\epsilon_{\ell},j}))$ for $j = 1,2, \ldots,M$;
	\item set $\paramvec_{\epsilon_{\ell-1}} = \left[k_{\epsilon_{\ell-1},1},k_{\epsilon_{\ell-1},2},\ldots,k_{\epsilon_{\ell-1},M}\right]$;
\end{enumerate}
Thus, the \emph{ABC multilevel Monte Carlo} (ABCMLMC)~\cite{Warne2018} method proceeds though computing the Monte Carlo estimate to \eqref{eq:ABCMLMC},
\begin{equation}
\label{eq:ABCMLMC_mc}
\hat{\paramvec}_{\epsilon_{\ell}} = \frac{1}{m_0}\sum_{i=1}^{m_0} \paramvec_{\epsilon_{0}}^{(i)} + \sum_{\ell=1}^{L}\frac{1}{m_\ell} \sum_{i=1}^{m_\ell}\left[\paramvec_{\epsilon_{\ell}}^{(i)} - \paramvec_{\epsilon_{\ell-1}}^{(i)}\right],
\end{equation}
where $\paramvec_{\epsilon_{0}}^{(1)},\ldots, \paramvec_{\epsilon_{0}}^{(m_0)}$ are $m_0$ independent samples using ABC rejection sampler and $(\paramvec_{\epsilon_{\ell}}^{(1)}, \paramvec_{\epsilon_{\ell-1}}^{(1)})$, $ \ldots,(\paramvec_{\epsilon_{\ell}}^{(m_\ell)}, \paramvec_{\epsilon_{\ell-1}}^{(m_\ell)})$ are $m_\ell$ independent sample pairs using the approximately correlated ABC rejection sampler for $\ell = 1,2, \ldots, L$. It is essential that each bias correction term in~\eqref{eq:ABCMLMC_mc} be estimated in ascending order to ensure the marginal CDF estimates for $\epsilon_{\ell-1}$ are available for use in the coupling scheme for the $\ell$th bias correction term. The complete algorithm is provided in Appendix~\ref{sec:app_ABCMLMC} and the example code is  provided in \texttt{ABCMLMC.m}. We refer the reader to Warne et~al.~\cite{Warne2018} for further details. 

Provided successive ABC posteriors are similar in \emph{correlation structure}, then the ABCMLMC method accelerates ABC rejection sampler through reduction of the number of $m_L$ samples required for a given error tolerance. 
However, additional bias is introduced by the approximately correlated ABC rejection sampler since the telescoping summation is not strictly satisfied. In practice, this affects the choice of the acceptance threshold sequence~\cite{Warne2018}. 

\begin{table}[h]
	\caption{Comparison of ABC methods for the mono-molecular inference problem. Estimates of the posterior mean are given along with the $95\%$ confidence intervals. Signed relative errors are shown for the confidence interval limits with respect to the true parameter values $k_1 = 1$, $k_2 = 0.1$, and $k_3 = 0.05$. Computations are performed using an Intel$^\text{\textregistered}$ Core\texttrademark~i7-5600U CPU (2.6 GHz).}	
	\begin{tabular}{l|lrr}
		Method & Posterior mean estimate & Relative error& Compute Time \\
		\hline
		 ABC rejection    & $\hat{k}_1 = 1.1690\times10^{0}\pm 7.1130\times10^{-2}$ & $[9.8\%,24.0\%]$ & $7,268$ (sec) \\ 
		 sampler & $\hat{k}_2 = 1.1011\times10^{-1}\pm4.5105\times10^{-3}$         & $[5.6\%,14.6\%]$&\\
		& $\hat{k}_3 = 5.3644\times10^{-2}\pm1.9500\times10^{-3}$         & $[3.4\%,11.2\%]$ &\\
		\hline
		ABCMCMC   & $\hat{k}_1 =  1.2786\times10^{0}\pm 7.8982\times10^{-2}$ & $[20.0\%,35.8\%]$& $8,059$ (sec) \\ 
		& $\hat{k}_2 = 1.1269\times10^{-1}\pm5.2836\times10^{-3}$            & $[7.4\%,18.0\%]$ & \\
		& $\hat{k}_3 = 5.5644\times10^{-2}\pm2.0561\times10^{-3}$            & $[7.2\%,15.4\%]$ &\\  
		\hline
		ABCSMC   & $\hat{k}_1 = 1.0444\times10^{0}\pm 6.4736\times10^{-2}$ & $[-2.0\%,10.9\%]$ & $580$ (sec) \\ 
		& $\hat{k}_2 = 9.8318\times10^{-2}\pm3.2527\times10^{-3}$ & $[-4.9\%,1.6\%]$& \\
		& $\hat{k}_3 = 4.8685\times10^{-2}\pm1.8032\times10^{-3}$ & $[-6.2\%,1.0\%]$&\\
		\hline
		ABCMLMC   & $\hat{k}_1 = 1.1535\times10^{0}\pm 4.8907\times10^{-2}$ & $[10.5\%,20.2\%]$ &$1,327$ (sec) \\ 
		& $\hat{k}_2 = 1.0654\times10^{-1}\pm4.8715\times10^{-3}$ & $[1.7\%,11.4\%]$ &\\
		& $\hat{k}_3 = 5.1265\times10^{-2}\pm2.2210\times10^{-3}$ & $[-2.0\%,7.0\%]$&\\            
	\end{tabular}
	\label{tab:abc_monomol}
\end{table}

We provide a comparison of the ABC rejection sampler, ABCMCMC, ABCSMC and ABCMLMC methods as presented here. Posterior means are computed for both the mono-molecular model and enzyme kinetics model inference problems along with the $95\%$ confidence intervals for each estimate\footnote{We report the 95\% confidence interval as an accuracy measure for the Monte Carlo estimate of the true posterior mean. This does not quantify parameter uncertainty that would be achieved through posterior covariances and credible intervals.} (see Appendix~\ref{sec:app_ABC_conf}). Computation times and parameter estimates are provided in Tables~\ref{tab:abc_monomol} and \ref{tab:abc_michment}. Example code, \texttt{DemoABCMethodsMonoMol.m} and \texttt{DemoABCMethodsMichMent.m}, is provided. Algorithm configurations are supplied in Appendix~\ref{sec:app_ABC_conf}. 

Note that no extensive tuning of the ABCMCMC, ABCSMC, or ABCMLMC algorithms is performed, thus these outcomes do not reflect a detailed and optimised implementation of the methods. Such a benchmark would be significantly more computationally intensive than the comparison of Monte Carlo methods for the forwards problem (Figure~\ref{fig:MLMC}). 
The computational statistics literature demonstrates that ABCMCMC~\cite{Marjoram2003}, ABCSMC~\cite{Beaumont2009,Sisson2007} and ABCMLMC~\cite{Guha2017,Jasra2017,Warne2018} can be tuned to provide very competitive results on a given inference problem. However, a large number of trial computations are often required to achieve this tuning, or more complex adaptive schemes need to be exploited~\cite{Drovandi2011,Roberts2009}. Instead, our comparisons represent a more practical guide in which computations are kept short and almost no tuning is performed, thus giving a fair comparison for a fixed, short computational budget.  

\begin{table}[h]
	\caption{Comparison of ABC methods for the enzyme kinetics model inference problem. Estimates of the posterior mean are given along with the $95\%$ confidence intervals. Signed relative errors are shown for the confidence interval limits with respect to the true parameter values $k_1 = 0.001$, $k_2 = 0.005$, and $k_3 = 0.01$. Computations are performed using an Intel$^\text{\textregistered}$ Core\texttrademark~i7-5600U CPU (2.6 GHz).}	
	\begin{tabular}{l|lrr}
		Method & Posterior mean estimate & Relative error & Compute time \\
		\hline
		 ABC rejection   & $\hat{k}_1 = 1.0098\times10^{-3}\pm 1.7011\times10^{-4}$ & $[-16.0\%,18.0\%]$ & $2,037$ (sec) \\ 
		 sampler & $\hat{k}_2 = 7.7203\times10^{-3}\pm7.3490\times10^{-4}$ & $[39.7\%,69.1\%]$& \\
		& $\hat{k}_3 = 1.5164\times10^{-2}\pm2.1201\times10^{-3}$ & $[30.4\%,72.8\%]$&\\
	    \hline
	    ABCMCMC   & $\hat{k}_1 = 3.0331\times10^{-4}\pm 7.5670\times10^{-5}$ & $[-77.2\%,-62.1\%]$ & $2,028$ (sec) \\ 
	    & $\hat{k}_2 = 5.8106\times10^{-3}\pm7.9993\times10^{-4}$ & $[0.2\%,32.2\%]$& \\
	    & $\hat{k}_3 = 3.3865\times10^{-2}\pm3.0187\times10^{-3}$ & $[208.5\%,268.8\%]$&\\
	    \hline
	    ABCSMC   & $\hat{k}_1 = 9.8972\times10^{-4}\pm 1.4630\times10^{-4}$ & $[-15.7\%,13.6\%]$ & $342$ (sec) \\ 
	    & $\hat{k}_2 = 9.2680\times10^{-3}\pm9.1997\times10^{-4}$ & $[67.0\%,103.8\%]$& \\
	    & $\hat{k}_3 = 1.3481\times10^{-2}\pm1.3278\times10^{-3}$ & $[21.5\%,48.1\%]$ &\\ 
	    \hline
	    ABCMLMC   & $\hat{k}_1 = 1.0859\times10^{-3}\pm 6.7058\times10^{-5}$ & $[1.9\%,15.3\%]$ & $795$ (sec) \\ 
	    & $\hat{k}_2 = 6.8111\times10^{-3}\pm3.2107\times10^{-4}$ & $[29.9\%,42.6\%]$ & \\
	    & $\hat{k}_3 = 1.3663\times10^{-2}\pm1.0729\times10^{-3}$ & $[25.9\%,47.4\%]$ &\\         
	\end{tabular}
	\label{tab:abc_michment}
\end{table}

For both inference problems, ABCSMC and ABCMLMC produce more accurate parameter estimates in less computation time than ABC rejection sampler and ABCMCMC. ABCSMC performs better on the mono-molecular chain model and ABCMLMC performs better on the enzyme kinetic model. This is not to say that ABCMCMC cannot be tuned to perform more efficiently, but it does indicate that ABCMCMC is harder to tune without more extensive testing. An advantage of ABCMLMC is that it has fewer components that require tuning, since the sample numbers may be optimally chosen, and the user need only provide a sequence of acceptance thresholds~\cite{Warne2018}. However, the ability to tune the proposal kernel in ABCSMC can be a significant advantage, despite the challenge of determining a good choice for such a proposal~\cite{Drovandi2011,Filippi2013}. 

\subsection{Summary of the inverse problem}
Bayesian approaches for uncertainty quantification and parameter inference have proven to be powerful techniques in the life sciences. However, in the study of intracellular biochemical processes, the intractability of the likelihood causes is a significant computational challenge.

A crucial advance towards obtaining numerical solutions in this Bayesian inverse problem setting has been the advent of likelihood-free methods that replace likelihood evaluations with stochastic simulation. Through such methods, especially those based on ABC, a direct connection between the forwards and inverse problems is made explicit. Not only is efficient forwards simulation key, but information obtained through each new sample path must be effectively used. While ABC rejection sampler ignores the latter, advanced methods like ABCMCMC, ABCSMC, and ABCMLMC all provide different mechanisms for incorporating this new information. For a specific inverse problem, however, it will often not be clear which method will perform optimally. In general, significant trial sampling must be performed to get the most out of any of these algorithms. 

The application of MLMC methods to ABC-based samplers is a very new area of research. There are many open questions that relate to the appropriateness of various approximations and coupling schemes for a given inference problem. The method of Warne et~al.~\cite{Warne2018} is conceptually straightforward, however, it is known that additional bias is incurred through the coupling scheme. Alternative coupling approximations that combine MLMC and MCMC~\cite{Guha2017} or SMC~\cite{Jasra2017} may be improvements, or a hybrid scheme could be derived through a combination of these methods. Currently, it is not clear which method is the most widely applicable.

\section{Summary and Outlook}

Stochastic models of biochemical reaction networks are routinely used to study of various intracellular processes, such as gene regulation. By studying the forwards problem, stochastic models can be explored \emph{in silico} to gain insight into potential hypotheses related to observed phenomena and inform potential experimentation. Models may be calibrated and key kinetic information may be extracted from time course data through the inverse problem. Both the forward and inverse problems are, in practice, reliant upon computation.

 
Throughout this review, we deliberately avoid detailed theoretical derivations in favour of practical computational or algorithmic examples. Since all of our code examples are specifically developed within the user friendly MATLAB$^\text{\textregistered}$ programming environment, our codes do not represent optimal implementations. Certainly, there are software packages available that implement some of these techniques and others that we have not discussed. For example, stochastic simulation is available in \texttt{COPASI}~\cite{Hoops2006}, \texttt{StochKit}~\cite{LiCao2007,Sanft2011}, \texttt{StochPy}~\cite{Maarleveld2013}, and \texttt{STEPS}~\cite{Hepburn2012}, and ABC-based inference is available in \texttt{ABC-SysBio}~\cite{Liepe2014}, \texttt{abctools}~\cite{Nunes2015}, and \texttt{ABCtoolbox}~\cite{Wegmann2010}. 

Throughout this review, we have focused on two biochemical reaction network models, the mono-molecular chain model for which the chemical master equation can  be solved analytically, and the enzyme kinetics model with an intractable chemical master equation. However, the majority of our MATLAB$^\text{\textregistered}$ implementations are completely general and apply to an arbitrary biochemical reaction network model. The biochemical reaction network construction functions, \texttt{MonoMolecularChain.m} and \texttt{MichaelisMenten.m}, demonstrate how to construct a biochemical reaction network data structure that most of our scripts and functions may directly use, with the only exceptions being those that relate to the analytical solution to the chemical master equation of the mono-molecular chain model. Therefore, our code may be viewed as a useful collection of prototype implementations that may be easily applied in new contexts.

The techniques we present here are powerful tools for the analysis of real biological data and design of experiments.
In particular, these methods are highly relevant for the reconstruction of genetic regulatory networks from gene expression microarray data~\cite{Cohen2015,Gao2018,Ocone2015}. Cell signalling pathways can also be analysed using mass-spectroscopy proteomic data~\cite{Wang2016,Tian2010}. Furthermore, Bayesian optimal experimental design using stochastic biochemical reaction networks is key to reliable characterisation of light-induced gene expression~\cite{Ruess2015}, a promising technique for external genetic regulation {\em in vivo}~\cite{Jayaraman2016,Yamada2018}.

This review provides researches in the life sciences with the fundamental concepts and computational tools required to apply stochastic biochemical reaction network models in practice. Through practical demonstration, the current state-of-the-art in simulation and Monte Carlo methods will be more widely and readily applied by the broader life sciences community.

\paragraph{Software availability}
The MATLAB$^\text{\textregistered}$ code examples and demonstration scripts are available from GitHub \href{https://github.com/ProfMJSimpson/Warne2018}{https://github.com/ProfMJSimpson/Warne2018}.
\paragraph{Acknowledgements}
This work was supported by the Australian Research Council (DP170100474). REB is a Royal Society Wolfson Research Merit Award holder, would like to thank the Leverhulme Trust for a Research Fellowship and also acknowledges the BBSRC for funding via grant no. BB/R000816/1. Computational resources where provided by the eResearch Office, Queensland University of Technology. We thank the two referees for helpful and insightful comments.

\begin{appendices}
	\appendix
	
	\section{Derivation of mono-molecular chain mean and variances using the chemical master equation}
	\label{sec:app_dev_cme}
	\numberwithin{equation}{section}
	\numberwithin{figure}{section}
	\numberwithin{algorithm}{section}
	\numberwithin{table}{section}
	\setcounter{equation}{0}
	In this section, we provide an example on how to derive moments of the chemical master equation (CME) solution without explicit CME evaluation. The presented analysis is specific to the two species mono-molecular chain model as presented in the main text. Our approach is based on the examples from Erban et~al.~\cite{Erban2007}, however, the result is more complex since we deal with a two chemical species, $A$ and $B$.
	
	For convenience, we restate the model. Here we consider a two species mono-molecular chain,
	\begin{equation}
		\label{eq:app_monomol}
		\emptyset \overset{k_1}{\rightarrow} A \overset{k_2}{\rightarrow} B \overset{k_3}{\rightarrow} \emptyset,
	\end{equation}
	with known kinetic rate parameters $k_1$, $k_2$ and $k_3$. Given the state vector, $\bvec{X}(t) = [A(t),B(t)]^T$, the respective propensity functions are
	\begin{equation}
		\label{eq:app_prop}
		a_1(\bvec{X}(t)) = k_1, \quad a_2(\bvec{X}(t)) = k_2 A(t), \quad a_3(\bvec{X}(t)) = k_3 B(t).
	\end{equation} The stoichiometric vectors are
	\begin{equation}
		\label{eq:app_nu}
		\nu_1 = \begin{bmatrix}
			1 \\
			0
		\end{bmatrix}, \quad \nu_2 = \begin{bmatrix}
			-1 \\
			1
		\end{bmatrix}, \quad \nu_3 = \begin{bmatrix}
			0 \\
			-1
		\end{bmatrix}.
	\end{equation}
	
	For $P(\bvec{x},t \mid \bvec{x}_0) = \CondProb{\bvec{X}(t) = \bvec{x}}{\bvec{X}(0) = \bvec{x}_0}$, the general form of the CME is
	\begin{equation}
		\label{eq:app_cme}
		\dydx{P(\bvec{x},t \mid \bvec{x}_0)}{t} = \sum_{j=1}^M a_j(\bvec{x} - \nu_j)P(\bvec{x} - \nu_j, t\mid \bvec{x}_0)- P(\bvec{x},t \mid \bvec{x}_0 ) \sum_{j=1}^M a_j(\bvec{x}).
	\end{equation} 
	After substituting the propensity functions (\eqref{eq:app_prop}) and stoichiometric vectors (\eqref{eq:app_nu}) into \eqref{eq:app_cme}, we obtain the CME specific to the mono-molecular chain model  (\eqref{eq:app_monomol})
	\begin{align}
		\label{eq:app_monomol_mce}
		\dydx{P(a,b,t\mid a_0,b_0)}{t} &= k_1 P(a-1,b,t\mid a_0,b_0) + k_2(a+1) P(a+1,b-1,t\mid a_0,b_0) \\ & \quad+ k_3(b+1) P(a,b+1,t\mid a_0,b_0) - (k_1 +k_2 a + k_3 b) P(a,b,t\mid a_0,b_0).\notag  
	\end{align}
	Henceforth, we will denote $p_{a,b}(t)$ as the solution to the mono-molecular CME (\eqref{eq:app_monomol_mce}).
	
	Rather than solve the full CME, we seek a solution to the mean copy number of $A$ at time $t$,
	\begin{equation}
		\label{eq:def_Ma}
		M_a(t) = \sum_{a=0}^{\infty} \sum_{b=0}^{\infty} a p_{a,b}(t),
	\end{equation}
	the mean copy number of $B$ at time $t$,
	\begin{equation}
		\label{eq:def_Mb}
		M_b(t) = \sum_{a=0}^{\infty} \sum_{b=0}^{\infty} b p_{a,b}(t),
	\end{equation}
	the variance of $A$ at time $t$,
	\begin{equation}
		\label{eq:def_Va}
		V_a(t) = \sum_{a=0}^{\infty} \sum_{b=0}^{\infty} \left(a - M_a(t)\right)^2 p_{a,b}(t),
	\end{equation}
	the variance of $B$ at time $t$,
	\begin{equation}
		\label{eq:def_Vb}
		V_b(t) = \sum_{a=0}^{\infty} \sum_{b=0}^{\infty} \left(b - M_b(t)\right)^2 p_{a,b}(t),
	\end{equation}
	and the covariance of $A$ and $B$ at time $t$,
	\begin{equation}
		\label{eq:def_Cab}
		C_{a,b}(t) = \sum_{a=0}^{\infty} \sum_{b=0}^{\infty} \left(a - M_a(t)\right)\left(b - M_b(t)\right) p_{a,b}(t).
	\end{equation}
	
	We will derive a system of ODEs that describe the evolution of $M_a(t)$,$M_b(t)$, $V_a(t)$, $V_b(t)$, and $C_{a,b}(t)$ without explicitly solving the CME in \eqref{eq:app_monomol_mce}. Instead we exploit the linearity of the derivative along with the property,
	\begin{equation}
		\label{eq:app_pme_prop}
		\sum_{a=0}^{\infty}\sum_{b=0}^{\infty} p_{a,b}(t) = 1,
	\end{equation}
	for all $t$.
	
	To derive an ODE for $M_a(t)$, we multiply \eqref{eq:app_monomol_mce} by $a$ and sum over all $a$ and $b$.
	\begin{align*}
		\dydx{}{t}\left[\sum_{a=0}^{\infty} \sum_{b=0}^{\infty} a p_{a,b}(t)\right] &= \sum_{a=1}^{\infty} \sum_{b=0}^{\infty} k_1a p_{a-1,b}(t) + \sum_{a=0}^{\infty} \sum_{b=1}^{\infty} k_2a(a+1) p_{a+1,b-1}(t) \\&\quad+ \sum_{a=0}^{\infty} \sum_{b=0}^{\infty} k_3a(b+1) p_{a,b+1}(t) - \sum_{a=0}^{\infty} \sum_{b=0}^{\infty} a (k_1 + k_2 a + k_3 b) p_{a,b}(t).
	\end{align*}
	After changing indices ($a-1 \to a$ in the first term, $a+1 \to a$ and $b-1 \to b$ in the second term, and $b+1 \to b$ in the third term),  we obtain 
	\begin{align*}
		\dydx{}{t}\left[\sum_{a=0}^{\infty} \sum_{b=0}^{\infty} a p_{a,b}(t)\right] &= \sum_{a=0}^{\infty} \sum_{b=0}^{\infty} k_1(a+1) p_{a,b}(t) + \sum_{a=0}^{\infty} \sum_{b=0}^{\infty} k_2(a-1)a p_{a,b}(t) \\&\quad+ \sum_{a=0}^{\infty} \sum_{b=0}^{\infty} k_3ab p_{a,b}(t) - \sum_{a=0}^{\infty} \sum_{b=0}^{\infty} a (k_1 + k_2 a + k_3 b) p_{a,b}(t).
	\end{align*}
	We simplify the right hand side,
	\begin{align*}
		\dydx{}{t}\left[\sum_{a=0}^{\infty} \sum_{b=0}^{\infty} a p_{a,b}(t)\right] &= k_1\sum_{a=0}^\infty \sum_{b=0}^\infty p_{a,b}(t) - k_2 \sum_{a=0}^\infty \sum_{b=0}^\infty a p_{a,b}(t), 
	\end{align*}
	and apply property (\ref{eq:app_pme_prop}) to give
	\begin{align*}
		\dydx{}{t}\left[\sum_{a=0}^{\infty} \sum_{b=0}^{\infty} a p_{a,b}(t)\right] &= k_1 - k_2 \sum_{a=0}^\infty \sum_{b=0}^\infty a p_{a,b}(t). 
	\end{align*}
	Using the definition of $M_a(t)$ (\eqref{eq:def_Ma}), we obtain the ODE for the mean of $A$,
	\begin{equation}
		\label{eq:app_Ma}
		\dydx{M_a(t)}{t} = k_1 - k_2 M_a(t).
	\end{equation}
	
	Similarly, we derive an ODE for $M_b(t)$ by muliplying \eqref{eq:app_monomol_mce} by $b$ and proceed in the same manner as we did for $M_a(t)$ 
	\begin{align*}
		\dydx{}{t}\left[\sum_{a=0}^{\infty} \sum_{b=0}^{\infty} b p_{a,b}(t)\right] &= \sum_{a=1}^{\infty} \sum_{b=0}^{\infty} k_1b p_{a-1,b}(t) + \sum_{a=0}^{\infty} \sum_{b=1}^{\infty} k_2b(a+1) p_{a+1,b-1}(t) \\&\quad+ \sum_{a=0}^{\infty} \sum_{b=0}^{\infty} k_3b(b+1) p_{a,b+1}(t) - \sum_{a=0}^{\infty} \sum_{b=0}^{\infty} b (k_1 + k_2 a + k_3 b) p_{a,b}(t) \\
		&= \sum_{a=0}^{\infty} \sum_{b=0}^{\infty} k_1b p_{a,b}(t) + \sum_{a=0}^{\infty} \sum_{b=0}^{\infty} k_2(b+1)a p_{a,b}(t) \\&\quad+ \sum_{a=0}^{\infty} \sum_{b=0}^{\infty} k_3(b-1)b p_{a,b}(t) - \sum_{a=0}^{\infty} \sum_{b=0}^{\infty} b (k_1 + k_2 a + k_3 b) p_{a,b}(t) \\
		&= k_2 \sum_{a=0}^\infty \sum_{b=0}^\infty a p_{a,b}(t) - k_3 \sum_{a=0}^\infty \sum_{b=0}^\infty b p_{a,b}(t).
	\end{align*}
	Using the definitions of $M_a(t)$ (\eqref{eq:def_Ma}) and $M_b(t)$ (\eqref{eq:app_Mb}) we obtain the ODE for the mean of $B$,
	\begin{equation}
		\label{eq:app_Mb}
		\dydx{M_b(t)}{t} = k_2 M_a(t) - k_3 M_b(t).
	\end{equation}
	
	To derive the ODE for $V_a(t)$, first note that through expanding \eqref{eq:def_Va} it can be shown that
	\begin{equation}
		\label{eq:VaMa_prop}
		V_a(t) + M_a(t)^2 = \sum_{a=0}^\infty \sum_{b=0}^\infty a^2  p_{a,b} (t).
	\end{equation}
	Thus, we multiply \eqref{eq:app_monomol_mce} by $a^2$, sum over all $a$ and $b$, change indices, and simplify as follows,
	\begin{align*}
		\dydx{}{t}\left[\sum_{a=0}^{\infty} \sum_{b=0}^{\infty} a^2 p_{a,b}(t)\right] &= \sum_{a=1}^{\infty} \sum_{b=0}^{\infty} k_1a^2 p_{a-1,b}(t) + \sum_{a=0}^{\infty} \sum_{b=1}^{\infty} k_2a^2(a+1) p_{a+1,b-1}(t) \\&\quad+ \sum_{a=0}^{\infty} \sum_{b=0}^{\infty} k_3a^2(b+1) p_{a,b+1}(t) - \sum_{a=0}^{\infty} \sum_{b=0}^{\infty} a^2 (k_1 + k_2 a + k_3 b) p_{a,b}(t) \\
		&= \sum_{a=0}^{\infty} \sum_{b=0}^{\infty} k_1(a+1)^2 p_{a,b}(t) + \sum_{a=0}^{\infty} \sum_{b=0}^{\infty} k_2(a-1)^2a p_{a,b}(t) \\&\quad+ \sum_{a=0}^{\infty} \sum_{b=0}^{\infty} k_3a^2b p_{a,b}(t) - \sum_{a=0}^{\infty} \sum_{b=0}^{\infty} a^2 (k_1 + k_2 a + k_3 b) p_{a,b}(t)\\
		&= k_1 \sum_{a=0}^\infty \sum_{b=0}^\infty p_{a,b}(t) +(2k_1 +k_2 )\sum_{a=0}^\infty \sum_{b=0}^\infty a p_{a,b}(t)  - 2k_2 \sum_{a=0}^\infty \sum_{b=0}^\infty a^2 p_{a,b}(t)\\
		&= k_1 + (2k_1+k_2) \sum_{a=0}^\infty \sum_{b=0}^\infty a p_{a,b}(t) - 2k_2 \sum_{a=0}^\infty \sum_{b=0}^\infty a^2 p_{a,b}(t).
	\end{align*} 
	Using the definition of $M_a(t)$ (\eqref{eq:app_Ma}) and property (\ref{eq:VaMa_prop}), we have the ODE,
	\begin{equation*}
		\dydx{}{t}\left[V_a(t) + M_a(t)^2\right] = k_1  + (2 k_1 + k_2)M_a(t) - 2k_2\left(V_a(t) + M_a(t)^2\right).
	\end{equation*}
	Apply the chain rule to obtain,
	\begin{equation}
		\label{eq:app_Va_part1}
		\dydx{V_a(t)}{t} = -2M_a(t)\dydx{M_a(t)}{t} + k_1  + (2 k_1 + k_2)M_a(t) - 2k_2\left(V_a(t) + M_a(t)^2\right),
	\end{equation}
	then substitute \eqref{eq:app_Ma} into \eqref{eq:app_Va_part1} and simplify to arrive at the ODE for the variance of $A$,
	\begin{equation}
		\label{eq:app_Va}
		\dydx{V_a(t)}{t} = k_1 + k_2M_a(t) - 2k_2 V_a(t).
	\end{equation}
	
	Similarly, to derive the ODE for $V_b(t)$ we note that through expanding \eqref{eq:def_Vb} and \eqref{eq:def_Cab} it can be shown that
	\begin{equation}
		\label{eq:VbMb_prop}
		V_b(t) + M_b(t)^2 = \sum_{a=0}^\infty \sum_{b=0}^\infty b^2  p_{a,b} (t),
	\end{equation}	
	and
	\begin{equation}
		\label{eq:CabMaMb_prop}
		C_{a,b}(t) + M_a(t)M_b(t) = \sum_{a=0}^\infty \sum_{b=0}^\infty ab p_{a,b} (t).
	\end{equation}
	Thus, we multiply \eqref{eq:app_monomol_mce} by $b^2$, sum over all $a$ and $b$, change indices, and simplify as follows,
	\begin{align*}
		\dydx{}{t}\left[\sum_{a=0}^{\infty} \sum_{b=0}^{\infty} b^2 p_{a,b}(t)\right] &= \sum_{a=1}^{\infty} \sum_{b=0}^{\infty} k_1b^2 p_{a-1,b}(t) + \sum_{a=0}^{\infty} \sum_{b=1}^{\infty} k_2b^2(a+1) p_{a+1,b-1}(t) \\&\quad+ \sum_{a=0}^{\infty} \sum_{b=0}^{\infty} k_3b^2(b+1) p_{a,b+1}(t) - \sum_{a=0}^{\infty} \sum_{b=0}^{\infty} b^2 (k_1 + k_2 a + k_3 b) p_{a,b}(t) \\
		&= \sum_{a=0}^{\infty} \sum_{b=0}^{\infty} k_1b^2 p_{a,b}(t) + \sum_{a=0}^{\infty} \sum_{b=0}^{\infty} k_2(b+1)^2a p_{a,b}(t) \\&\quad+ \sum_{a=0}^{\infty} \sum_{b=0}^{\infty} k_3(b-1)^2b p_{a,b}(t) - \sum_{a=0}^{\infty} \sum_{b=0}^{\infty} b^2 (k_1 + k_2 a + k_3 b) p_{a,b}(t)\\
		&= k_2 \sum_{a=0}^{\infty} \sum_{b=0}^{\infty} a p_{a,b}(t) + k_3 \sum_{a=0}^{\infty} \sum_{b=0}^{\infty} b p_{a,b}(t) \\ 
		&\quad+ 2k_2\sum_{a=0}^{\infty} \sum_{b=0}^{\infty} ab p_{a,b}(t) - 2k_3\sum_{a=0}^{\infty} \sum_{b=0}^{\infty} b^2 p_{a,b}(t).
	\end{align*}
	Using the definitions of $M_a(t)$ (\eqref{eq:def_Ma}) and $M_b(t)$ (\eqref{eq:def_Mb}), and properties (\ref{eq:VbMb_prop}) and (\ref{eq:CabMaMb_prop}) we have the ODE
	\begin{equation*}
		\dydx{}{t}\left[V_b(t) + M_b(t)^2\right] = k_2 M_a(t) + k_3 M_b(t) + 2k_2\left(C_{a,b}(t) + M_a(t)M_b(t)\right) -2 k_3\left(V_b(t) + M_b(t)^2\right).
	\end{equation*}
	We apply the chain rule
	\begin{align}
		\label{eq:app_Vb_part1}
		\dydx{V_b(t)}{t} &= -2M_b(t)\dydx{M_b(t)}{t}+ k_2 M_a(t) + k_3 M_b(t) \\&\quad+ 2k_2\left(C_{a,b}(t) + M_a(t)M_b(t)\right) -2 k_3\left(V_b(t) + M_b(t)^2\right),\notag
	\end{align}
	then substitute \eqref{eq:app_Mb} into \eqref{eq:app_Vb_part1} and simplify to obtain the ODE for the variance of $B$
	\begin{equation}
		\label{eq:app_Vb}
		\dydx{V_b(t)}{t} = k_2 M_a(t) + k_3 M_b(t) +2 k_2 C_{a,b} - 2k_3 V_b(t).
	\end{equation}

	Finally, we derive the ODE for $C_{a,b}(t)$ by multiplying \eqref{eq:app_monomol_mce} by $ab$, summing over all $a$ and $b$, changing indices, and simplifying as follows:
	\begin{align*}
		\dydx{}{t}\left[\sum_{a=0}^{\infty} \sum_{b=0}^{\infty} ab p_{a,b}(t)\right] &= \sum_{a=1}^{\infty} \sum_{b=0}^{\infty} k_1 ab p_{a-1,b}(t) + \sum_{a=0}^{\infty} \sum_{b=1}^{\infty} k_2ab(a+1) p_{a+1,b-1}(t) \\&\quad+ \sum_{a=0}^{\infty} \sum_{b=0}^{\infty} k_3ab(b+1) p_{a,b+1}(t) - \sum_{a=0}^{\infty} \sum_{b=0}^{\infty} ab (k_1 + k_2 a + k_3 b) p_{a,b}(t) \\
		&= \sum_{a=0}^{\infty} \sum_{b=0}^{\infty} k_1 (a+1)b p_{a,b}(t) + \sum_{a=0}^{\infty} \sum_{b=0}^{\infty} k_2(a-1)(b+1)a p_{a,b}(t) \\&\quad+ \sum_{a=0}^{\infty} \sum_{b=0}^{\infty} k_3a(b-1)b p_{a,b}(t) - \sum_{a=0}^{\infty} \sum_{b=0}^{\infty} ab (k_1 + k_2 a + k_3 b) p_{a,b}(t) \\
		&= k_1 \sum_{a=0}^{\infty} \sum_{b=0}^{\infty} b p_{a,b}(t) - k_2 \sum_{a=0}^{\infty} \sum_{b=0}^{\infty} b p_{a,b}(t) \\
		&\quad -(k_2+k_3)\sum_{a=0}^{\infty} \sum_{b=0}^{\infty} ab p_{a,b}(t) + k_2 \sum_{a=0}^{\infty} \sum_{b=0}^{\infty} a^2 p_{a,b}(t). 
	\end{align*}
	Using the definition of $M_a(t)$ (\eqref{eq:def_Ma}) and $M_b(t)$ (\eqref{eq:def_Mb}), and properties~(\ref{eq:VaMa_prop}) and (\ref{eq:CabMaMb_prop}) we obtain the ODE
	\begin{align*}
		\dydx{}{t}\left[C_{a,b}(t) + M_a(t)M_b(t)\right] &= k_1 M_a(t) - k_2M_b(t) -(k_2+k_3) \left(C_{a,b}(t) + M_a(t)M_b(t)\right)\\  &\quad + k_2 \left(V_a(t) + M_a(t)^2\right).
	\end{align*}
	Apply the chain rule and product rule
	\begin{align}
		\label{eq:app_Cab_part1}
		\dydx{C_{a,b}(t)}{t} &=  - M_a(t)\dydx{M_b(t)}{t} - M_b(t)\dydx{M_a(t)}{t} + k_1 M_a(t) - k_2M_b(t)\\  &\quad -(k_2+k_3) \left(C_{a,b}(t) + M_a(t)M_b(t)\right) + k_2 \left(V_a(t) + M_a(t)^2\right),\notag
	\end{align}
	then substitute \eqref{eq:app_Ma} and \eqref{eq:app_Mb} into \eqref{eq:app_Cab_part1} and simplify to obtain the ODE for the covariance of $A$ and $B$
	\begin{equation}
		\label{eq:app_Cab}
		\dydx{C_{a,b}(t)}{t} = k_2 V_a(t) - k_2 M_a(t) - (k_2 + k_3) C_{a,b}(t).
	\end{equation}
	
	Therefore, Equations~(\ref{eq:app_Ma}),(\ref{eq:app_Mb}),(\ref{eq:app_Va}),(\ref{eq:app_Vb}), and (\ref{eq:app_Cab}) form a non-homogeneous linear system of ODEs,
	\begin{align*}
		\dydx{M_a(t)}{t} &= k_1 - k_2 M_a(t), \\ 
		\dydx{M_b(t)}{t} &= k_2 M_a(t) - k_3 M_b (t), \\
		\dydx{V_a(t)}{t} &= k_1 + k_2 M_a(t) - 2k_2 V_a(t), \\ 
		\dydx{V_b(t)}{t} &= k_2 M_a(t) + k_3 M_b(t) + 2 k_2 C_{a,b}(t) - k_3V_b(t), \\
		\dydx{C_{a,b}(t)}{t} &= k_2 V_a(t) - k_2 M_a(t) - (k_2 + k_3)C_{a,b}(t).
	\end{align*}
	After solving for the homogeneous solution, a particular solution may be obtained through using the method of undetermined coefficients. Given the initial conditions $A(0) = a_0$  and $B(0) = b_0$ with probability one, the solution, in the case when $k_2 \neq k_3$, is
	\begin{align}
		M_a(t) &= \frac{k_1}{k_2} + \left(a_0 - \frac{k_1}{k_2}\right)e^{-k_2 t}, \label{eq:app_MaSol}\\
		M_b(t) &=  \frac{k_1}{k_3} + \frac{k_2 a_0 - k_1}{k_3 - k_2}e^{-k_2 t} +\left(b_0 - \frac{k_2 a_0 - k_1}{k_3 - k_2} - \frac{k_1}{k_3}\right)e^{-k_3 t}, \label{eq:app_MbSol}\\
		V_a(t) &= \frac{k_1}{k_2} +\left(a_0 - \frac{k_1}{k_2}\right)e^{-k_2 t} -a_0 e^{-2k_2 t}, \label{eq:app_VaSol}\\
		V_b(t) &= \frac{k_1}{k_3} + \frac{k_2 a_0 - k_1}{k_3 - k_2}e^{-k_2 t} + \left(b_0 - \frac{k_2 a_0 - k_1}{k_3 - k_2} - \frac{k_1}{k_3}\right)e^{-k_3 t} + \frac{2a_0k_2^2}{k_3^2 - k_2^2}e^{-(k_3 + k_2)t} \notag\\ &\quad - \frac{a_0 k_2}{k_3 - k_2}e^{-2k_2 t} + \left[\frac{a_0k_2}{k_3 - k_2}\left(1 - \frac{2k_2}{k_3 + k_2}\right) - b_0\right]e^{-2k_3 t}, \label{eq:app_VbSol}\\
		C_{a,b}(t) &= -\frac{a_0k_2}{k_3 - k_2}e^{-(k_3 + k_2)t} + \frac{a_0k_2}{k_3 - k_2}e^{-2k_2t}. \label{eq:app_CabSol}
	\end{align}
	Equations~(\ref{eq:app_MaSol})--(\ref{eq:app_CabSol}) are the time dependent solutions for the moments and from these solutions we can evaluate the long time limit, $t \to \infty$, to give simple expression for the associated stationary solutions,
	\begin{align*}
		\lim_{t \to \infty} M_a(t) &= \frac{k_1}{k_2}, \\ \lim_{t \to \infty} V_a(t) &= \frac{k_1}{k_2}, \\
		\lim_{t \to \infty} M_b(t) &= \frac{k_1}{k_3}, \\ \lim_{t \to \infty} V_b(t) &= \frac{k_1}{k_3}, \\
		\lim_{t \to \infty} C_{a,b}(t) &= 0.
	\end{align*}
	See the example codes \texttt{DemoCMEMeanVar.m} and \texttt{DemoStationaryDist.m} for the evaluation of this solution.
	
	\newpage
	
	\section{Evaluation of the mono-molecular chain chemical master equation solution}
	\label{sec:app_cme_comp}
	Jahnke and Huisinga~\cite{Jahnke2007} derive an analytic solution to the CME for a general mono-molecular BCRN. Applying their general solution to the species mono-molecular chain (\eqref{eq:app_monomol}) results in the following general solution to the CME (\eqref{eq:app_monomol_mce}),
	\begin{equation}
		\label{eq:monomol_cme_sol}
		P(a,b,t \mid a_0, b_0) = \mathcal{P}\left(a,b,\lambda_a(t),\lambda_b(t)\right) \ast \mathcal{M}\left(a,b,a_0,\alpha_a(t),\alpha_b(t)\right) \ast \mathcal{M}\left(a,b,b_0,\beta_a(t),\beta_b(t)\right),
	\end{equation}
	where $\ast$ is the discrete convolution operation~\cite{Jahnke2007}. $\mathcal{P}\left(a,b,\lambda_a(t),\lambda_b(t)\right)$ is a product Poisson distribution, given by
	\begin{equation}
		\label{eq:def_prodpois}
		\mathcal{P}\left(a,b,\lambda_a(t),\lambda_b(t)\right) =\begin{cases} 
			\dfrac{\lambda_a(t)^a}{a!} \dfrac{\lambda_b(t)^b}{b!} e^{-(|\lambda_a(t)| + |\lambda_b(t)|)}, &\, \text{if } a \geq 0, b \geq 0, \\
			0 &\, \text{otherwise},
		\end{cases}
	\end{equation}
	where the functions $\lambda_a(t)$ and $\lambda_b(t)$ are obtained through the initial value problem (IVP)
	\begin{align}
		\label{eq:ivplambda}
		\dydx{\lambda_a(t)}{t} = k_1 - k_2 \lambda_a(t),\quad \dydx{\lambda_b(t)}{t} = k_2\lambda_a(t) - k_3 \lambda_b(t), \quad t > 0,
	\end{align}
	with initial conditions $\lambda_a(0) = \lambda_b(0) = 0$.$\mathcal{M}\left(a,b,a_0,\alpha_a(t),\alpha_b(t)\right)$ and  $\mathcal{M}\left(a,b,b_0,\beta_a(t),\beta_b(t)\right)$ are multinomial distributions, given by
	\begin{equation}
		\label{eq:def_multa}
		\mathcal{M}\left(a,b,a_0,\alpha_a(t),\alpha_b(t)\right) =\begin{cases} 
			a_0!\dfrac{\left(1-|\alpha_a(t)|-|\alpha_b(t)|\right)^{a_0 - |a|-|b|}}{(a_0 - |a| - |b|)!}\dfrac{\alpha_a(t)^a}{a!}\dfrac{\alpha_b(t)^b}{b!}  &\, \text{if } |a| + |b| \leq a_0, \\
			0 &\, \text{otherwise},
		\end{cases}
	\end{equation}
	and
	\begin{equation}
		\label{eq:def_multb}
		\mathcal{M}\left(a,b,b_0,\beta_a(t),\beta_b(t)\right) =\begin{cases} 
			b_0!\dfrac{\left(1-|\beta_a(t)|-|\beta_b(t)|\right)^{b_0 - |a|-|b|}}{(b_0 - |a| - |b|)!}\dfrac{\beta_a(t)^a}{a!}\dfrac{\beta_b(t)^b}{b!}  &\, \text{if } |a| + |b| \leq b_0, \\
			0 &\, \text{otherwise}.
		\end{cases}
	\end{equation} 
	The functions $\alpha_a(t)$, $\alpha_b(t)$, $\beta_a(t)$ and $\beta_b(t)$ are obtained through the IVPs
	\begin{align}
		\label{eq:ivpalpha}
		\dydx{\alpha_a(t)}{t} = -k_2 \alpha_a(t), \quad \dydx{\alpha_b(t)}{t} = k_2 \alpha_a(t)  - k_3 \alpha_b(t), \quad t > 0 ,
	\end{align}
	and
	\begin{align}
		\label{eq:ivpbeta}
		\dydx{\beta_a(t)}{t} = -k_2 \beta_a(t), \quad \dydx{\beta_b(t)}{t} = k_2 \beta_a(t)  - k_3 \beta_b(t), \quad t > 0 ,
	\end{align}
	with initial conditions $\alpha_a(0) = 1$, $\alpha_b(0) = 0$, $\beta_a(0) = 0$, and $\beta_b(0) = 1$.
	
	\eqref{eq:monomol_cme_sol} represents a direct substitution of the two species mono-molecular chain into the general solution by Jahnke and Huisinga~\cite{Jahnke2007}. However, direct point-wise evaluation of this solution is not feasible. Specifically, there are two challenges: (i) the two convolutions are taken over an infinite two-dimensional integer lattice; and (ii) the non-zero probabilities in the product Poisson and Multinomial distribution can be so small that numerical underflow/overflow is almost certain. The first issue can be solved by determining the finite set of lattice sites that do not contribute to the convolutions, this can be achieved by invoking specific features of \eqref{eq:app_monomol}. The second issue requires that we perform calculations using logarithms of probabilities rather than the true probabilities. Extra care must be taken in the convolution summations.
	
	We first simplify the convolution operations to ensure finite computations. Solving the IVPs \ref{eq:ivplambda},\ref{eq:ivpalpha}, and \ref{eq:ivpbeta} yields, for $k_2 \neq k_3$,
	\begin{align}
		\lambda_a(t) &=\frac{k_1}{k_2}\left(1 - e^{-k_2 t}\right), \label{eq:lambda_a}\\
		\lambda_b(t) &= \frac{k_1}{k_3} + \frac{k_1}{k_3 - k_2}\left[e^{-k_2 t} + (k_2 - 2k_3)e^{-k_3 t}\right],\label{eq:lambda_b}\\
		\alpha_a(t) &= e^{-k_2 t},\label{eq:alpha_a}\\
		\alpha_b(t) &= \frac{k_2}{k_3 - k_2}\left(e^{-k_2 t} - e^{-k_3 t}\right),\label{eq:alpha_b}\\
		\beta_a(t) &= 0,\label{eq:beta_a}\\
		\beta_b(t) &= e^{-k_3 t} \label{eq:beta_b}.
	\end{align}   
	A key result is that $\beta_a(t)$ is zero for all time (\eqref{eq:beta_a}). Through substitution of \eqref{eq:beta_a} into \eqref{eq:def_multb}, we have 
	\begin{equation}
		\label{eq:def_multb_2}
		\mathcal{M}\left(a,b,b_0,0,\beta_b(t)\right) =\begin{cases} 
			b_0!\dfrac{\left(1-|\beta_b(t)|\right)^{b_0 - |a|-|b|}}{(b_0 - |a| - |b|)!}\dfrac{0^a}{a!}\dfrac{\beta_b(t)^b}{b!}  &\, \text{if } |a| + |b| \leq b_0, \\
			0 &\, \text{otherwise}.
		\end{cases}
	\end{equation}
	This implies that $\mathcal{M}\left(a,b,b_0,\beta_a(t),\beta_b(t)\right) = 0$ if $a \neq 0$. That is,
	\begin{equation}
		\label{eq:def_multb_3}
		\mathcal{M}\left(a,b,b_0,0,\beta_b(t)\right) =\begin{cases} 
			b_0!\dfrac{\left(1-|\beta_b(t)|\right)^{b_0 -|b|}}{(b_0 - |b|)!}\dfrac{\beta_b(t)^b}{b!}  &\, \text{if } a = 0, \text{ and } |b| \leq b_0, \\
			0 &\, \text{otherwise}.
		\end{cases}
	\end{equation}
	
	We can now make a significant simplification of the second convolution in~\eqref{eq:monomol_cme_sol}. Let $\mathcal{M}_a(a,b,t) = \mathcal{M}\left(a,b,a_0,\alpha_a(t),\alpha_b(t)\right)$ and $\mathcal{M}_b(a,b,t) = \mathcal{M}\left(a,b,b_0,0,\beta_b(t)\right)$, we have
	\begin{align*}
		\mathcal{M}_a(a,b,t) \ast \mathcal{M}_b(a,b,t) &= \sum_{a_w \in \mathbb{N}} \sum_{b_w \in \mathbb{N}}  \mathcal{M}_a(a_w,b_w,t)\mathcal{M}_b(a-a_w,b-b_w,t) \\
		&= \sum_{b_w \in \mathbb{N}} \mathcal{M}_a(a,b_w,t)\mathcal{M}_b(0,b-b_w,t),
	\end{align*}
	where $\mathbb{N} = \mathbb{Z}^+ \cup \{0\}$.
	By \eqref{eq:def_multb_3}, $\mathcal{M}_b(0,b,t) = 0$ if $|b| > b_0$. Furthermore, we have $b \geq 0$ from the definition of the BCRN (\eqref{eq:app_monomol}). It follows that only terms with $b \geq b_w \geq \max(0,b-b_0)$ can contribute to the convolution, that is, 
	\begin{align*}
		\mathcal{M}_a(a,b,t) \ast \mathcal{M}_b(a,b,t) = \sum_{b_w = \max(0,b-b_0)}^{b} \mathcal{M}_a(a,b_w,t)\mathcal{M}_b(0,b-b_w,t).
	\end{align*}
	While this convolution never involves more that $b$ terms, we can apply a further constraint on the upper bound of the index. By \eqref{eq:def_multa} we have $\mathcal{M}_a(a,b,t) = 0$ if $|a| + |b| > a_0$. Since $a \geq 0$ and $b \geq 0$ from the definition of the BCRN (\eqref{eq:app_monomol}. That is, terms with $b_w \geq a_0 - a \geq 0$ will not contribute to the convolution. Therefore, the multinomial convolution term in \eqref{eq:monomol_cme_sol} is 
	\begin{equation}
		\label{eq:conv_mult}
		\mathcal{M}_a(a,b,t) \ast \mathcal{M}_b(a,b,t) = \sum_{b_w = \max(0,b-b_0)}^{\min(b,\max(0,a_0 - a))} \mathcal{M}_a(a,b_w,t)\mathcal{M}_b(0,b-b_w,t).
	\end{equation}
	
	Let $\mathcal{P}(a,b,t) = \mathcal{P}\left(a,b,\lambda_a(t),\lambda_b(t)\right)$ and substitute~\eqref{eq:conv_mult} into \eqref{eq:monomol_cme_sol} to yield
	\begin{align*}
		P(a,b,t \mid a_0, b_0) &= \mathcal{P}\left(a,b,t\right) \ast \left[\sum_{b_w = \max(0,b-b_0)}^{\min(b,\max(0,a_0 - a))} \mathcal{M}_a(a,b_w,t)\mathcal{M}_b(0,b-b_w,t)\right] \\
		&= \sum_{a_z \in \mathbb{N}} \sum_{b_z \in \mathbb{N}} \mathcal{P}\left(a-a_z,b-b_z,t\right)\left[\sum_{b_w = \max(0,b_z-b_0)}^{\min(b_z,\max(0,a_0 - a_z))} \mathcal{M}_a(a_z,b_w,t)\mathcal{M}_b(0,b_z-b_w,t)\right].
	\end{align*}
	By definition of the product Poisson distribution (\eqref{eq:def_prodpois}), $\mathcal{P}\left(a,b,t\right) = 0$ for $a < 0$ or $b < 0$. Hence, only terms with $a \geq a_z$ and $b \geq b_z$ contribute to the convolution. Therefore, we obtain the following expression for \eqref{eq:monomol_cme_sol}
	\begin{align}
		\label{eq:monomol_cme_sol_nz}
		P(a,b,t \mid a_0, b_0) &= \sum_{a_z=0}^{a} \sum_{b_z=0}^b \mathcal{P}\left(a-a_z,b-b_z,t\right)\\ 
		&\quad\quad \times\left[\sum_{b_w = \max(0,b_z-b_0)}^{\min(b_z,\max(0,a_0 - a_z))} \mathcal{M}_a(a_z,b_w,t)\mathcal{M}_b(0,b_z-b_w,t)\right], \notag
	\end{align} 
	which requires $\mathcal{O}({ab^2})$ evaluations of either \eqref{eq:def_prodpois}, \eqref{eq:def_multa} or \eqref{eq:def_multb_3}.
	
	Now that we have bounded the number of operations required to evaluate the solution of the CME, we now address the problem of numerical overflow/underflow. There are two possible sources for this type of numerical error. Firstly, the factorials and products of powers involved in the evaluation of \eqref{eq:def_prodpois}, \eqref{eq:def_multa}, and \eqref{eq:def_multb_3} can be very large, causing overflow. Secondly, the probabilities in the convolution terms can be very small, causing underflow.  
	
	To avoid these issues we work with logarithms of probabilities. For the non-zero cases of Equations (\ref{eq:def_prodpois}), (\ref{eq:def_multa}) and (\ref{eq:def_multb_3}), we have 
	\begin{align}
		\ln \mathcal{P}\left(a,b,t\right) &= a\ln \lambda_a(t) + b\ln \lambda_b(t) -(|\lambda_a(t)| + |\lambda_b(t)|) - \sum_{a_i = 1}^{a} \ln a_i -\sum_{b_i = 1}^{b} \ln b_i, \label{eq:lnPoi}\\
		\ln \mathcal{M}_a\left(a,b,t\right) &= a\ln \alpha_a(t) + b\ln \alpha_b(t) + (a_0 - a - b)\ln\left(1 - \alpha_a(t) - \alpha_b(t)\right)\notag\\ &\quad+ \sum_{a_i =a_0 -a -b}^{a_0} \ln a_i - \sum_{a_i =1}^a \ln a_i - \sum_{b_i =1}^b \ln b_i,\label{eq:lnMula}\\                                    
		\ln \mathcal{M}_b\left(a,b,t\right) &= b\ln \beta_b(t) + (b_0 - b)\ln\left(1 - \beta_b(t)\right) +\sum_{b_i =b_0 -b}^{b_0} \ln b_i - \sum_{b_i =1}^b \ln b_i. \label{eq:lnMulb}
	\end{align}
	Equations ~(\ref{eq:lnPoi})--(\ref{eq:lnMulb}) enable the computation to proceed with overflow or underflow being significantly less likely. Therefore, we take the logarithm of \eqref{eq:monomol_cme_sol_nz} to obtain
	\begin{equation}
		\label{eq:lnP}
		\ln P(a,b,t \mid a_0, b_0) = \ln\left[\sum_{a_z=0}^{a} \sum_{b_z=0}^b e^{\ln \mathcal{P}\left(a-a_z,b-b_z,t\right) + \ln \mathcal{F}\left(a_z,b_z,t\right)}\right],
	\end{equation} 
	where
	\begin{equation}
		\label{eq:lnF}
		\ln \mathcal{F}(a_z,b_z,t) = \ln\left[\sum_{b_w = \max(0,b_z-b_0)}^{\min(b_z,\max(0,a_0 - a_z))} e^{\ln\mathcal{M}_a(a_z,b_w,t) + \ln\mathcal{M}_b(0,b_z-b_w,t)}\right].
	\end{equation}
	Computing the logarithms of summations of exponential functions in \eqref{eq:lnP} and \eqref{eq:lnF} is still prone to overflow and underflow since the probabilities will be very small in practice. A common solution to numerically stable logarithm of summations of exponential functions is known as the ``log-sum-exp trick''. This works by noting, for any $x,y\in \mathbb{R}$, that
	\begin{align*}
		\ln\left[e^x + e^y\right] &= \ln\left[\left(e^{x-\max(x,y)} + e^{y-\max(x,y)}\right) e^{\max(x,y)}\right]\\
		&= \ln\left[e^{x-\max(x,y)} + e^{y-\max(x,y)}\right] + \max(x,y).
	\end{align*}
	Thus, computations are re-scaled to the natural scale of the terms in the summation, thus terms that do underflow would not have affected the result significantly. Now, let
	\begin{equation*}
		R(a_z,b_z) = \max_{b_w \in \left[\max(0,b_z-b_0),\min(b_z,\max(0,a_0 - a_z))\right]}\left\{\ln\mathcal{M}_a(a_z,b_w,t) + \ln\mathcal{M}_b(0,b_z-b_w,t)\right\},
	\end{equation*}
	and 
	\begin{equation*}
		S(a,b) = \max_{[a_z,b_z] \in \left[0,a\right]\times\left[0,b\right]}\left\{\ln \mathcal{P}\left(a-a_z,b-b_z,t\right) + \ln \mathcal{F}\left(a_z,b_z,t\right)\right\}.
	\end{equation*}
	Then use $S(a,b)$ and $R(a_z,b_z)$ with the ``log-sum-exp trick'' to yield a numerically robust form of \eqref{eq:lnP}. That is,
	\begin{equation}
		\label{eq:lnPlse}
		\ln P(a,b,t \mid a_0, b_0) = \ln\left[\sum_{a_z=0}^{a} \sum_{b_z=0}^b e^{\ln \mathcal{P}\left(a-a_z,b-b_z,t\right) + \ln \mathcal{F}\left(a_z,b_z,t\right)-S(a,b)}\right] + S(a,b),
	\end{equation} 
	where
	\begin{equation}
		\label{eq:lnFlse}
		\ln \mathcal{F}(a_z,b_z,t) = \ln\left[\sum_{b_w = \max(0,b_z-b_0)}^{\min(b_z,\max(0,a_0 - a_z))} e^{\ln\mathcal{M}_a(a_z,b_w,t) + \ln\mathcal{M}_b(0,b_z-b_w,t)-R(a_z,b_z)}\right] + R(a_z,b_z).
	\end{equation} 
	The example code, \texttt{CMEsolMonoMol.m} provides a numerical implementation of the CME solution (\eqref{eq:monomol_cme_sol}) using \eqref{eq:lnPlse} and \eqref{eq:lnFlse}.
	
	\newpage
	
	\section{Synthetic data}
	\label{sec:app_data}
	The synthetic data used in the main manuscript and example code is provide in Table~\ref{tab:monomoldat} for the mono-molecular chain model and in Table~\ref{tab:michmentdat} for the enzyme kinetics model.
	
	\begin{table}[h]
		\centering
		\caption{Data, $\bvec{Y}_{\text{obs}}$,  used for inference on the mono-molecular chain. Generated using true parameter values $k_1 = 1.0$, $k_2 = 0.1$, and $k_3 = 0.05$ and initial conditions, $A(0) = 100$ and $B(0)= 0$. }	
		\begin{tabular}{l|rrrr}
			& $\bvec{Y}(t_1)$ & $\bvec{Y}(t_2)$ & $\bvec{Y}(t_3)$ & $\bvec{Y}(t_4)$ \\
			\hline
			$t$ & $25$ & $50$ & $75$ & $100$ \\
			$A(t)$ & $14$ & $12$ & $17$ & $15$ \\
			$B(t)$ & $68$ & $34$ & $14$ & $14$
		\end{tabular}
		\label{tab:monomoldat}
	\end{table}
	
	\begin{table}[h]
		\centering
		\caption{Data, $\bvec{Y}_{\text{obs}}$,  used for inference on the enzyme kinetic model. Generated using true parameter values $k_1 = 0.001$, $k_2 = 0.005$, and $k_3 = 0.01$ and initial conditions $S(0) = E(0) = 100$ and $C(0)= P(0) = 0$. }	
		\begin{tabular}{l|rrrrr}
			& $\bvec{Y}(t_1)$ & $\bvec{Y}(t_2)$ & $\bvec{Y}(t_3)$ & $\bvec{Y}(t_4)$ & $\bvec{Y}(t_5)$\\
			\hline
			$t$ & $0$& $20$ & $40$ & $60$ & $80$ \\
			$P(t)$ & $0$ & $5$ & $16$ & $28$ & $39$ \\
			$P(t) + \mathbf{\xi}$ & $2.04$ & $6.99$ & $14.30$ & $28.71$ & $38.14$ \\
		\end{tabular}
		\label{tab:michmentdat}
	\end{table}
	
	\newpage
	
	\section{Additional ABC results}
	\label{sec:app_results}
	In the main manuscript only marginal probability densities are used to demonstrate ABC convergence. Here we present plot matrices with the bivariate marginals also. 
	
\begin{figure}
		\centering
		\includegraphics[width=\linewidth]{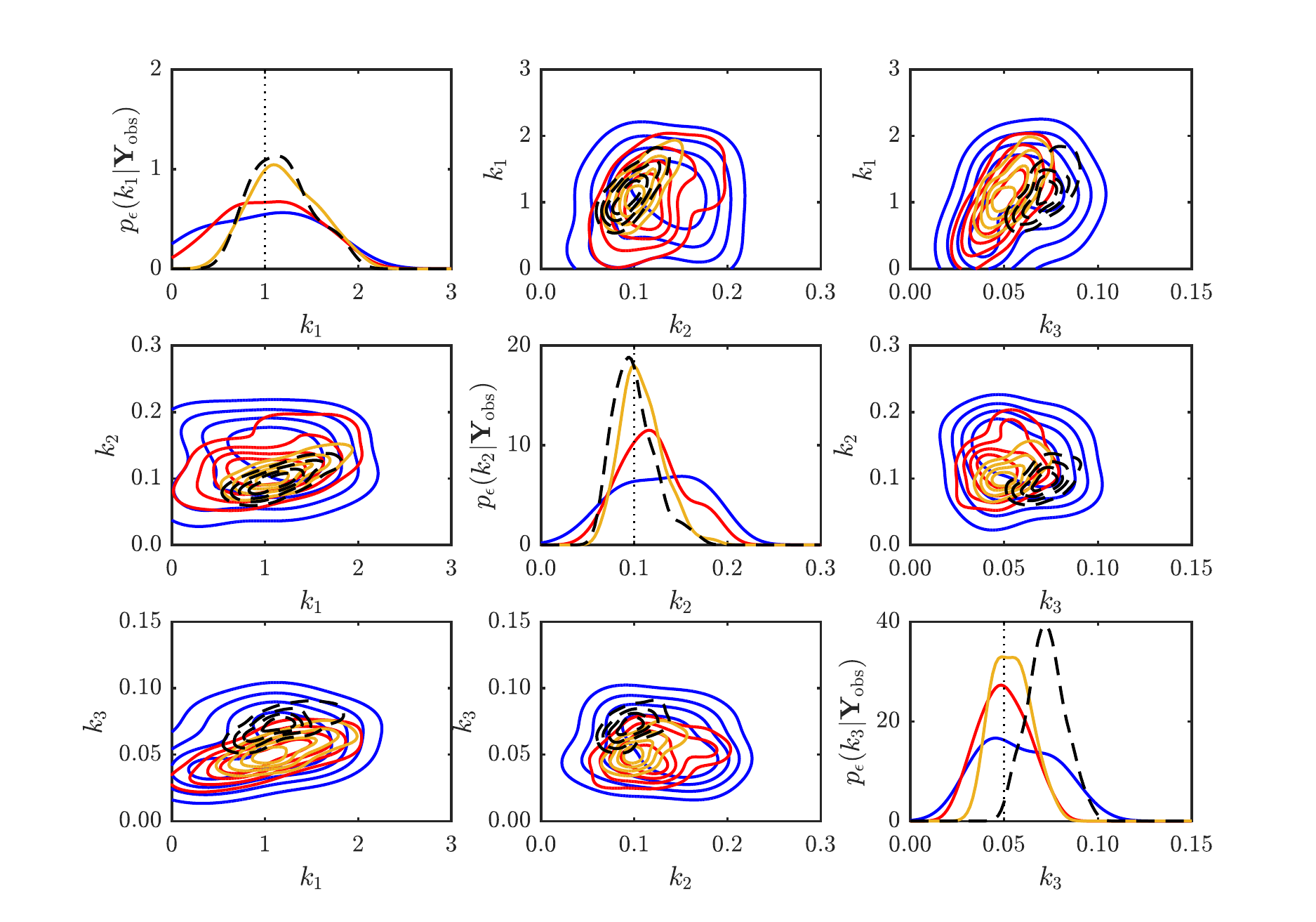}
		\caption{Convergence of ABC posterior to the true posterior as $\epsilon \to 0$ for the mono-molecular chain inference problem. Marginal posteriors are plotted for $\epsilon = 50$ (blue solid), $\epsilon = 25$ (red solid), $\epsilon = 12.5$ (yellow solid), and $\epsilon = 0$ (black dashed). Here, the $\epsilon = 0$ case corresponds to the exact likelihoods using the CME solution. Univariate marginals are plotted on the diagonals and bivariate marginals on off diagonal elements. Contour lines in bivariate marginal plots are selected such that six equal probability density intervals are  shown.  The true parameter values (black dotted) are $k_1 = 1.0$, $k_2 = 0.1$ and $k_3 = 0.05$. Note that the exact Bayesian posterior does not recover the true parameter  for $k_3$.}
		\label{fig:fig1sm}
	\end{figure}

	Since the mono-molecular chain model has three rate parameters, we have three univariate marginals posteriors and three bivariate marginal posteriors.
	Through application of the ABC with acceptance threshold, $\epsilon$, the equivalent marginals are
	\begin{align}
		\CondPDF{k_1}{\discrep{\bvec{Y}_{\text{obs}}}{\bvec{S}_{\text{obs}}} \leq \epsilon} &= \iint_{\mathbb{R}^2} \CondPDF{k_1,k_2,k_3}{\discrep{\bvec{Y}_{\text{obs}}}{\bvec{S}_{\text{obs}}} \leq \epsilon}\, \text{d}k_2\, \text{d}k_3, \label{eq:umar1} \\ 
		\CondPDF{k_2}{\discrep{\bvec{Y}_{\text{obs}}}{\bvec{S}_{\text{obs}}} \leq \epsilon} &= \iint_{\mathbb{R}^2} \CondPDF{k_1,k_2,k_3}{\discrep{\bvec{Y}_{\text{obs}}}{\bvec{S}_{\text{obs}}} \leq \epsilon}\, \text{d}k_1\, \text{d}k_3, \label{eq:umar2} \\
		\CondPDF{k_3}{\discrep{\bvec{Y}_{\text{obs}}}{\bvec{S}_{\text{obs}}} \leq \epsilon} &= \iint_{\mathbb{R}^2} \CondPDF{k_1,k_2,k_3}{\discrep{\bvec{Y}_{\text{obs}}}{\bvec{S}_{\text{obs}}} \leq \epsilon}\, \text{d}k_1\, \text{d}k_2, \label{eq:umar3}\\
		\CondPDF{k_1,k_2}{\discrep{\bvec{Y}_{\text{obs}}}{\bvec{S}_{\text{obs}}} \leq \epsilon} &= \int_{\mathbb{R}} \CondPDF{k_1,k_2,k_3}{\discrep{\bvec{Y}_{\text{obs}}}{\bvec{S}_{\text{obs}}} \leq \epsilon}\, \text{d}k_3, \label{eq:bmar1}\\ 
		\CondPDF{k_1,k_3}{\discrep{\bvec{Y}_{\text{obs}}}{\bvec{S}_{\text{obs}}} \leq \epsilon} &= \int_{\mathbb{R}} \CondPDF{k_1,k_2,k_3}{\discrep{\bvec{Y}_{\text{obs}}}{\bvec{S}_{\text{obs}}} \leq \epsilon}\, \text{d}k_2, \label{eq:bmar2}\\
		\CondPDF{k_2,k_3}{\discrep{\bvec{Y}_{\text{obs}}}{\bvec{S}_{\text{obs}}} \leq \epsilon} &= \int_{\mathbb{R}} \CondPDF{k_1,k_2,k_3}{\discrep{\bvec{Y}_{\text{obs}}}{\bvec{S}_{\text{obs}}} \leq \epsilon}\, \text{d}k_1. \label{eq:bmar3} 
	\end{align}
	The exact univariate and bivariate marginal posteriors are plotted against the ABC posterior for $\epsilon = [50,25,12.5,0]$ (with $\epsilon = 0$ meaning the exact posterior is sampled using the CME-based likelihood). Equations~(\ref{eq:umar1})--(\ref{eq:bmar3}) are plotted in Figure~\ref{fig:fig1sm}.
	
\begin{figure}
		\centering
		\includegraphics[width=0.55\linewidth]{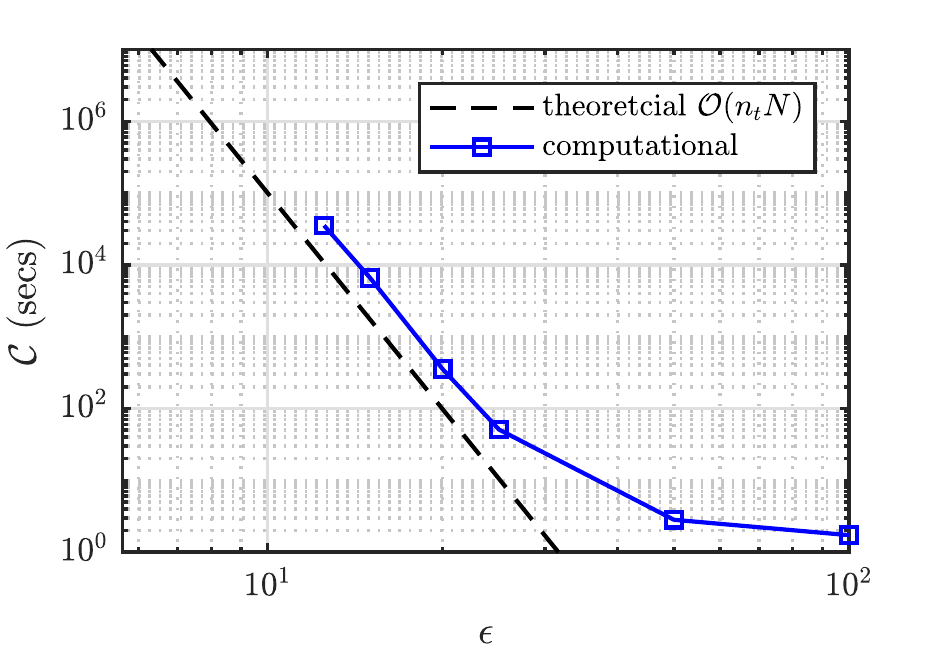}
		\caption{Computation time growth of ABC rejection sampling against theoretical result. Computations are performed using an Intel$^\text{\textregistered}$ Core\texttrademark~i7-5600U CPU (2.6 GHz).}
		\label{fig:fig2sm}
	\end{figure}

	Reducing $\epsilon$ further than $12.5$ is prohibitive, even for the mono-molecular chain model. Both Barber et~al.~\cite{Barber2015} and Fearnhead and Prangle~\cite{Fearnhead2012} provide an asymptotic result for the computation time, $\mathcal{C}$, as a function of $\epsilon$, that is, $\mathcal{C} = \mathcal{O}(\epsilon^{-d})$, where $d$ is the dimensionality of the data used in the ABC inference. For the synthetic data we have from Table~\ref{tab:monomoldat}, we have $d = n_t N$. Figure~\ref{fig:fig2sm} demonstrates that the computation times we obtain are consistent with this. 
	
	\newpage
	
	\section{ABC Multilevel Monte Carlo}
	\label{sec:app_ABCMLMC}
	Here we provide the ABC Multilevel Monte Carlo scheme (ABCMLMC) (see Warne~et~al.~\cite{Warne2018} for more details and the derivation). This particular implementation computes an estimate of the posterior mean.
	Given a sequence of acceptance thresholds, $\epsilon_0 > \epsilon_1 > \cdots > \epsilon_L = \epsilon$, and a sequence of sample numbers $m_0 > m_1 > \cdots > m_L$ (see Giles~\cite{Giles2008} and Warne et al.~\cite{Warne2018} for details on optimally computing the sample numbers), ABCMLMC proceeds as follows:
	\begin{enumerate}
		\item initialise $\ell = 0$;
		\item set $i = 1$\label{line:startlevel};
		\item generate a prior sample $\paramvec^* \sim \PDF{\paramvec}$\label{line:samplepriorABC2};
		\item generate simulated data, $\bvec{S}_{\text{obs}}^* \sim s(\bvec{S}_{\text{obs}}; \paramvec^*)$;
		\item if $\discrep{\bvec{Y}_{\text{obs}}}{\bvec{S}_{\text{obs}}^*} \leq \epsilon_\ell$, accept $\paramvec_{\epsilon_\ell}^{(i)} = \paramvec^*$ and set $i = i+1$, otherwise continue;
		\item if $i \leq m_\ell$,  go to step~\ref{line:samplepriorABC2}, otherwise continue;
		\item set $\hat{F}_{\ell,j}(z) = \sum_{i=1}^{m_\ell}\ind{z}{k_{\epsilon_\ell,j}^{(i)}} / m_\ell$ for $j = 1,2,\ldots,M$;
		\item if $\ell = 0$, then set $\hat{\paramvec_\epsilon} = \sum_{i=1}^{m_\ell}\paramvec_{\epsilon_\ell}^{(i)} / m_\ell$, set $\ell = \ell + 1$, and go to step~\ref{line:startlevel}, otherwise continue;
		\item set $i = 1$\label{line:startcoupling};
		\item set $k_{\epsilon_{\ell-1},j}^{(i)} = \hat{F}^{-1}_{\ell-1,j}(\hat{F}_{\ell,j}(k_{\epsilon_\ell,j}^{(i)}))$ for $j = 1,2,\ldots, M$;
		\item set $\paramvec_{\epsilon_{\ell-1}}^{(i)} = [k_{\epsilon_{\ell-1},1}^{(i)},k_{\epsilon_{\ell-1},2}^{(i)},\ldots,k_{\epsilon_{\ell-1},M}^{(i)}]$ and set $i = i+1$;
		\item if $i \leq m_\ell$, then go to step~\ref{line:startcoupling}, otherwise continue;
		\item set $\hat{F}_{\ell,j}(z) = \hat{F}_{\ell-1,j}(z) + \sum_{i=1}^{m_\ell} \left(\ind{z}{k_{\epsilon_\ell,j}^{(i)}} - \ind{z}{k_{\epsilon_{\ell-1},j}^{(i)}}\right)/m_\ell$;
		\item set $\hat{\paramvec_\epsilon} = \hat{\paramvec_\epsilon}  + \sum_{i=1}^{m_\ell}\left(\paramvec_{\epsilon_\ell}^{(i)} - \paramvec_{\epsilon_{\ell-j}}^{(i)}\right) / m_\ell$;
		\item if $\ell = L$,  then terminate, otherwise set $\ell = \ell + 1$ and go to step~\ref{line:startlevel};
	\end{enumerate} 
	
	\newpage
	
	\section{ABC algorithm configurations and additional results}
	\label{sec:app_ABC_conf}
	The following algorithm configurations for the ABC rejection sampler, ABCMCMC, ABCSMC and ABCMLMC are used to generate the results in Tables~3 and 4 of the main manuscript. The parameters are also contained in the code examples, \texttt{DemoABCMethodsMonoMol.m} and \texttt{DemoABCMethodsMichMent.m} .
	
	For the mono-molecular chain model inference problem each algorithm is configured as follows: for the ABC rejection sampler we set $m = 100$ and $\epsilon = 15$; for ABCMCMC we set $m_n = 500,000$, $m_b = 100,000$, $m_h = 10,000$, the proposal kernel is a Gaussian random walk with covariance matrix, $\Sigma = \diag(1\times 10^{-3},1\times 10^{-5}, 2.5\times 10^{-5})$, and $\epsilon = 15$; for ABCSMC we use $m_p = 100$, the proposal kernel  is a Gaussian random walk with covariance matrix, $\Sigma = \diag(1\times 10^{-3},1\times 10^{-5}, 2.5\times 10^{-5})$, and the discrepancy threshold sequence is $\epsilon_1 = 100$ with $\epsilon_{r+1} = \epsilon_r/2$ for $r = 2,3, \ldots, 5$; for ABCMLMC we use the discrepancy threshold sequence $\epsilon_0 = 100$ with $\epsilon_{\ell+1} = \epsilon_\ell/2$ for $\ell = 1,2, \ldots, 4$, and the sample number sequence, $m_0 = 800$, with $m_{\ell+1} = M_\ell/2$ for $\ell = 1,2, \ldots, 4$. For the prior, we assume all parameters are independent of each other and uniformly distributed with $k_1 \sim \mathcal{U}(0,2)$, $k_2 \sim \mathcal{U}(0,0.2)$, and $k_3 \sim \mathcal{U}(0,0.1)$.

	Similarly for the enzyme kinetics model inference problem each algorithm is configured as follows: for the ABC rejection sampler we set $m = 100$ and $\epsilon = 2.5$; for ABCMCMC we set $m_n = 500,000$, $m_b = 100,000$, $m_h = 10,000$, the proposal kernel is a Gaussian random walk with covariance matrix, $\Sigma = \diag(2.25\times 10^{-8},5.625\times 10^{-7}, 6.25\times 10^{-6})$, and $\epsilon = 2.5$; for ABCSMC we use $m_p = 100$, the proposal kernel  is a Gaussian random walk with covariance matrix, $\Sigma = \diag(2.25\times 10^{-8},5.625\times 10^{-7}, 6.25\times 10^{-6})$, and the discrepancy threshold sequence is $\epsilon_1 = 40$ with $\epsilon_{r+1} = \epsilon_r/2$ for $r = 2,3, \ldots, 5$; for ABCMLMC we use the discrepancy threshold sequence $\epsilon_0 = 40$ with $\epsilon_{\ell+1} = \epsilon_\ell/2$ for $\ell = 1,2, \ldots, 4$, and the sample number sequence, $m_0 = 800$, with $m_{\ell+1} = M_\ell/2$ for $\ell = 1,2, \ldots, 4$. For the prior, we assume all parameters are independent of each other and uniformly distributed with $k_1 \sim \mathcal{U}(0,0.003)$, $k_2 \sim \mathcal{U}(0,0.015)$, and $k_3 \sim \mathcal{U}(0,0.05)$.

\begin{figure}[h]
		\centering
		\includegraphics[width=\linewidth]{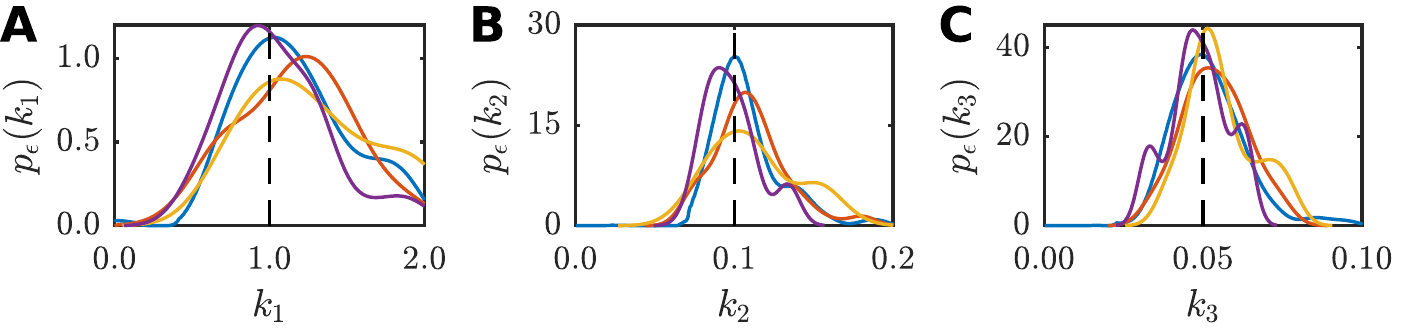}
		\caption{Comparison of ABC posteriors generated by the ABC rejection sampler (red solid), ABCMCMC (yellow solid), ABCSMC (purple solid) and ABCMLMC (blue solid) for the mono-molecular chain inference problem. The true parameter values (black dashed) are $k_1 = 1.0$, $k_2 = 0.1$ and $k_3 = 0.05$.}
		\label{fig:fig3sm}
	\end{figure}

	The resulting marginal posterior distributions are presented in Figure~\ref{fig:fig3sm}. ABCSMC and ABCMLMC recover the true parameters effectively and as less computationally intensive. 
	For the enzyme kinetic inference problem, more tuning and samples are required to obtain good estimates of the full marginal posterior distributions. Especially, since the ABCMCMC trajectory undergoes a long excursion into the low density tails.

\end{appendices}


\end{document}